\documentclass[twocolumn]{aastex62}

\usepackage{xcolor}
\newcommand{\nnb}[1]{\textcolor{black}{{#1}}}

\received{\today}

\shorttitle{Sample article}
\shortauthors{Barczynski et al.}

\usepackage{bm}
\usepackage{graphicx}
\graphicspath{{./}{figures/}{figs/}}
\usepackage{natbib}
\usepackage{amsmath}
\usepackage{esint}
\usepackage{hyperref}
\hypersetup{
    colorlinks=true,
    linkcolor=blue,
    filecolor=magenta,      
    urlcolor=cyan,
}
\usepackage{amsmath}
\usepackage{booktabs}
\usepackage{gensymb}
\usepackage{scalerel}
\usepackage{soul}

\begin{document}

\title{Electric current evolution at the footpoints of solar eruptions.}

\correspondingauthor{Krzysztof Barczynski}
\email{krzysztof.barczynski@pmodwrc.ch}

\author[0000-0001-7090-6180]{Krzysztof Barczynski}
\affil{LESIA, Observatoire de Paris, Universit\'e PSL , CNRS, Sorbonne Universit\'e, Universit\'e Paris-Diderot, 5 place Jules Janssen, 92190 Meudon, France}
\affil{Physikalisch Meteorologisches Observatorium Davos, World Radiation Center, Davos, Switzerland}
\affil{ Department of Physics, Eidgen\"ossische Technische Hochschule (ETH) Z\"urich, Z\"urich, Switzerland}

\author[0000-0001-5810-1566]{Guillaume Aulanier}
\affil{LESIA, Observatoire de Paris, Universit\'e PSL , CNRS, Sorbonne Universit\'e, Universit\'e Paris-Diderot, 5 place Jules Janssen, 92190 Meudon, France}

\author[0000-0002-6203-5239]{Miho Janvier}
\affil{Institut d'Astrophysique Spatiale, CNRS, Universit\'e Paris-Sud, Universit\'e Paris-Saclay, B\^at 121, 91405 Orsay CEDEX, France}

\author[0000-0003-3364-9183]{Brigitte Schmieder}
\affil{LESIA, Observatoire de Paris, Universit\'e PSL , CNRS, Sorbonne Universit\'e, Universit\'e Paris-Diderot, 5 place Jules Janssen, 92190 Meudon, France}

\author[0000-0002-6376-1144]{Sophie Masson}
\affil{LESIA, Observatoire de Paris, Universit\'e PSL , CNRS, Sorbonne Universit\'e, Universit\'e Paris-Diderot, 5 place Jules Janssen, 92190 Meudon, France}



\begin{abstract}
Electric currents play a critical role in the triggering  of solar flares and their evolution.
The aim of the present  paper is to test whether the surface electric current has a surface or subsurface fixed source as predicts the circuit approach of flare physics, or  is the response of the surface magnetic field to the evolution of the coronal magnetic field as the MHD approach proposes.
\nnb{Out of all 19 X-class flares as observed by SDO from 2011 to 2016 near the disk center, we analyzed the only 9 eruptive flares for which clear ribbon-hooks were identifiable.} Flare ribbons with hooks are considered to be the footprints of eruptive flux ropes in MHD flare models. For the first time, fine measurements of time-evolution of electric currents inside the   hooks in the observations  as well as   in the OHM  3D  MHD simulation are performed.
Our analysis shows a decrease of the electric current in the area surrounded by the ribbon hooks during and after the eruption. 
We interpret the decrease of the electric currents as due to the expansion of the flux rope in the corona during the eruption.
 Our analysis  brings a \nnb{new contribution} to the standard flare model in 3D.

\end{abstract}


\section{Introduction} \label{sec:intro}


       Solar flares and coronal mass ejections (CMEs) are the 
most energetic events of the active Sun \citep[][]{EmslieEtal2012,
SchrijverEtal2012}. They also constitute the strongest drivers of 
space weather \citep[][]{CliverDietrich2013,SchmiederAulanier2018a}. 
At their source regions in the low corona and within sunspots, 
the sub-Alfvenic plasma velocities and the low plasma $\beta$ altogether 
have two major implications. The first is that solar eruptions must 
draw their energy from current-carrying magnetic fields $\bm{B}$. 
The second is that preeruptive currents are almost colinear with 
the magnetic field, hence in a quasi force-free state. These 
properties constitute the base of all existing eruption models 
\citep[see e.g.][for reviews]{AlfvenCarlqvist1967,VantendKuperus1978,
Spicer1982,Forbes2000,Vrsnak2008,Aulanier2014,JanvierEtal2015}. 

Characterizing coronal currents $\bm{j}$ before and 
during eruptions is therefore important to develop a comprehensive 
understanding of the triggering and acceleration of CMEs, especially in the context of a growing need for accurate and quantitative space 
weather forecasts. 
In addition, on a more fundamental basis, specifying the 
spatio-temporal evolution of eruption-related currents at the Sun's 
surface would also bring an unprecedented input to the $\bm{v};\bm{B}$ 
vs $\bm{E};\bm{j}$ debate \citep[e.g.][]{Melrose1991,
Parker1996,Parker2001,Heikkila1997,Vasyliunas2005}, in particular 
regarding \nnb{whether the magnetic field} or the electric current 
should be considered as a prime variable and boundary condition 
in solar eruption models in particular, and in space plasmas in 
general. 
All these considerations recently led to a renewed interest 
in studying solar electric currents \citep[see e.g.][]{Melrose2017,
Georgoulis2018,SchmiederAulanier2018b}.

The view on the spatial distribution of currents in solar active regions has evolved over the years. For a long time, the geometrical 
properties on cylindrical flux ropes naturally led to consider that active region magnetic fields ought to be shielded from their 
surroundings through external return currents, with a magnitude equal to that of the direct currents. This so-called 
current neutralization of coronal flux-ropes was considered to 
be a natural prediction of the $\bm{v};\bm{B}$ paradigm, 
in the context of the MHD twisting of coronal flux-tubes \citep[as explained e.g. in][]{Parker1996,DalmasseEtal2015}. 
Such neutralized currents in the corona, however, are 
problematic for several eruption models that rely on the existence 
of net currents \citep[e.g.][]{VantendKuperus1978,ForbesPriest1995,
KliemTorok2006,DemoulinAulanier2010}. In addition, the most recent 
observations of sunspots revealed that many flaring active regions carry a non-neutralized (i.e. a net) current \citep[][]{RavindraEtal2011,
GeorgoulisEtal2012,ZhaoEtal2016,ChengDing2016,Vemareddy2019}. These 
observations could have been considered to support the 
$\bm{E};\bm{j}$ paradigm, with some circuit models associated with net coronal electric currents directly originating from the depths of the Sun's interior \citep[][]{AlfvenCarlqvist1967,Melrose1991}. 
But recent numerical MHD models of magnetic fields in the corona driven by magnetic flux emergence \citep[][]{TorokEtal2014} or by line-tied 
motions \citep[][]{DalmasseEtal2015} have both unveiled that MHD 
processes could also generate net coronal currents in pre-eruptive 
active regions, due to the shear component of the magnetic field across polarity investion lines (PIL). Two different analytical interpretations were 
provided to explain this association, either by invoking Lorentz 
forces \citep[][]{GeorgoulisEtal2012} or a simple property of 
the integral form of Amp\`ere's law with force-free fields 
\citep[][]{DalmasseEtal2015}. In both cases, it was shown 
that the distribution of net currents in solar active regions was 
also compatible with the $\bm{v};\bm{B}$ paradigm.

While the time-evolution of electric currents during solar eruptions is much less understood, it is worth noting that the different physical paradigms provide different predictions for it. 
On one hand, the $\bm{v};\bm{B}$ paradigm states that the 
current is a secondary variable. So its magnitude and variability 
should be locally constrained by the magnetic field gradients, 
following Amp\`ere's law \citep[as physically argued e.g. by][]
{Vasyliunas2005}. In this case, the right photospheric 
boundary-conditions of MHD coronal models should be line-tied. 
On the other hand, the $\bm{E};\bm{j}$ paradigm states that 
the current itself is the fundamental quantity. So coronal currents 
should originate from the Sun's interior \citep[as argued e.g. 
by][]{Melrose1991}. In this case, the current should be regarded 
as the relevant boundary condition for coronal-models. And in 
turn the magnetic field should be globally constrained by the 
distribution of currents following the integral version of 
Amp\`ere's equation, i.e. the Biot-Savart law.
Characterizing the evolution of electric currents at the 
footpoints of erupting flux-ropes could thus provide a novel way 
to test these two clear-cut and opposite predictions. One promising 
approach relies on the forward modeling of off-limb measurements 
within coronal cavities \citep[][]{DalmasseEtal2019}. Another 
approach relies on the investigation of whether the CME expansion 
during an eruption has any observable effect on the photospheric 
magnetic fields and electric currents at the footpoints of 
erupting flux ropes. 
There have already been some reports of magnetic feedbacks 
at the Sun's surface being induced by solar flares and CMEs. One 
example is the increase of horizontal-fields around PILs 
\citep{HudsonEtal2008, 2013SoPh..287..415P, 2017ApJ...839...67S, 2019ApJ...877...67B}. 
Another example is the amplification of narrow electric-currents 
inside spreading flare ribbons \citep[][]{2014ApJ...788...60J,
2016A&A...591A.141J, 2015A&A...580A.106M,SharykinEtal2019}, and possibly 
at the source region of sunquakes \citep[][]{SharykinKosovichev2015}.

The idea of measuring the time-evolution of currents at the footpoints of erupting flux-ropes  comes to us (see the review of   \citet{SchmiederAulanier2018b})
with the pioneer observational paper of 
\citet{ChengDing2016}.
From  the vector magnetic-field measurements  made during four flares in different active regions, the latter authors  reported that the direct computed electric current at all the (flux rope) footpoints with a strong enough magnetic field experiences a decrease  and proposed that this decrease was related to the decrease of the twist per unit length  imposed by the conservation of the total twist.
The Amp\`ere's law and the field-line equation imply that the increasing length $L_z$ of a flux rope leading to $B_\phi \propto 1/L_z$ and naturally leads to a decrease of axial current-densities $\jmath_z$.
So the observational result of \citet{ChengDing2016} tends to favor the 
predictions of the $\bm{v};\bm{B}$ paradigm.

To the author's knowledge, the only analysis of the
time-evolution of currents at the footpoints of erupting flux-ropes
from vector magnetic-field measurements was performed in the
pioneering work of \citet{ChengDing2016}, as also discussed
in the review by \citet{SchmiederAulanier2018b}. From the analysis
of four flares, \citet{ChengDing2016} reported that ``the direct current
at all the footpoints with stronger magnetic fields experience a decrease''.
And they proposed that this decrease was related to the fact that ``due
to total twist conservation, the twist per unit length decreases''. The
latter claim relies on considering the Amp\`ere's law, that governs the
axial current along a cylindrical flux tube as a function of its length and
twist, and on considering an increasing length and a fixed end-to-end twist,
as predicted by line-tied MHD. So these measurements tend to favor the
predictions of the MHD paradigm. But this conclusion is not certain yet.
Firstly because the currents measured by \citet{ChengDing2016} decreased
by $\simeq 7-13 \%$ only. This is not very much for a expanding flux rope
within a CME. Yet this may instead simply reflect the noise level of these
demanding measurements. Secondly because the exact surface-areas of
erupting flux-ropes are not obvious to identify. Considering areas that
roughly surround the endpoints of sigmoidal EUV loops as done by
\citet{ChengDing2016} may very well be sufficient. But these areas may also comprize unrelated currents, and therefore blur the signal. In short,
while the unprecedented results of \citet{ChengDing2016} favor the
$\vec{v};\vec{B}$ paradigm over the $\vec{E};\vec{j}$ approach, they
still need to be confirmed and fine-tuned with independent methods.

Performing fine-tuned measurements of the time-evolution of 
electric currents at the well-defined footpoints of erupting flux-ropes 
in an MHD model (Section~\ref{sec:simulation}) and in a series of observed eruptions (Sections~\ref{sec:obser_and_prep_data} and \ref{sec:electric_current_obs_foot}) is thus the main goal of the present paper. In order to identify the flux-rope footpoint-areas 
as accurately as possible in the considered MHD model as well as in the observations, 
we will consider \nnb{four} key topological properties of flux ropes.

Firstly, flux ropes embedded in coronal arcades are associated 
with quasi-separatrix layers (QSLs), which footprints display 
a double-J shape, and which curved endings constitute so-called 
hooks that sharply surround the flux-rope footpoints \citep[][]
{DemoulinPriestLonie1996,Titov2007,PariatDemoulin2012}. Secondly, 
the hooks (as well as the rest) of double-J shaped QSLs are 
observable during eruptions as bright flare-ribbons \citep[in 
particular in warm EUV-channels, see e.g.][]{2016A&A...591A.141J,
ZhaoEtal2016,Savcheva2016} as well as strong and narrow 
electric current concentrations \citep[see e.g.][]{JanvierEtal2013,
2014ApJ...788...60J, 2019A&A...621A..72A}. Thirdly, the QSL-hooks (and 
therefore their associated ribbons, currents, and \nnb{some} flux-rope 
footpoints) \nnb{drift in space and change shape in the course of the eruption  as due to series of reconnections as seen in numerical models \citep{2019A&A...621A..72A} and in SDO observations \citep{ZemanovaEtal2019, 2019ApJ...887..118C, 2019ApJ...887...71D, 2019ApJ...885...83L}.}
\nnb{Fourthly, flux ropes can also reconnect with large-scale solar arcades and jump to distant locations \citep{2010JGRA..11510104C, 2011ApJ...738..127L}}.

       These properties being considered altogether allow to 
\nnb{\it{select footpoint areas that belong to and remain within 
erupting flux ropes}} during the peak and main phase of all 
studied eruptions.
\nnb{So this paper focuses on the evolution of electric currents at the footpoints of non-reconnecting eruptive flux-rope field-lines.}

\section{Simulation}\label{sec:simulation}

\subsection{Description of simulation}\label{sec:simulation_descr}
Numerical simulations allow us to study the intrinsic mechanisms of the evolution of an eruptive flare that can then be compared with their observations at the Sun.
In our work, we analyzed a 3D MHD flare simulation provided by \citet{2015ApJ...814..126Z}.
For this simulation, the OHM-MPI code \citep{2005A&A...444..961A} is used to solve the zero-$\beta$ (pressureless), time-dependent 3D MHD equations which reproduce the flux rope expansion initiated by a torus-unstable magnetic structure.
The simulation uses a nonuniform mesh area $n_{x}\times n_{y} \times n_{z} = 251 \times 251 \times 231$ that covers the physical domain $x,y\in [-10; 10]$ and $z\in[0; 30]$ \citep{2015ApJ...814..126Z}.
The simulation provides the spatial and temporal evolution of the vector magnetic field ($\bm{B}$), mass density ($\rho$), and plasma velocity ($\bm{u}$).
The output of the simulation is presented in dimensionless unit, where the space-time unit $L=1$ is the distance between the PIL and the center of one magnetic field polarity at $z=t=0$; the time-unit $t_A=1$ is the propagating time of the Alfv\'en waves from the PIL to the center of the one magnetic field polarity; the magnetic permeability is set to $\mu_0=1$.
The results in dimensionless units can be scaled to the physical value \citep[][Section 2]{2019ApJ...877...67B}.

Using the simulation labelled ``Run D2" \citep{2015ApJ...814..126Z}, we studied an idealized bipolar active region with two asymmetric magnetic field concentrations. 
We focused on the limited domain of $x\in[-3;  2]$,  $y\in[-4.5; 3]$, and $z\in[0; 5]$ that cover the whole flare region.
We analyzed the temporal evolution from $t_{0}=164t_{A}$ right before the eruption onset ($t=165t_{A}$) to the end of simulation at $t_{end}=244t_{A}$.

\subsection{Ribbons and hook}
We derived the current density vector ($\bm{j}$) from the vector magnetic field ($\bm{B}$) obtained from the 3D MHD simulation to Amp\`ere circuital law,
\begin{equation} \label{eq:ampere_law}
\mu_{0}\bm{j} = \bm{\nabla} \bm{\times} \bm{B}.
\end{equation} 
In our analysis, we used the centered difference method for each mesh point.

\begin{figure*}[ht!]
\epsscale{1.05}
\plotone{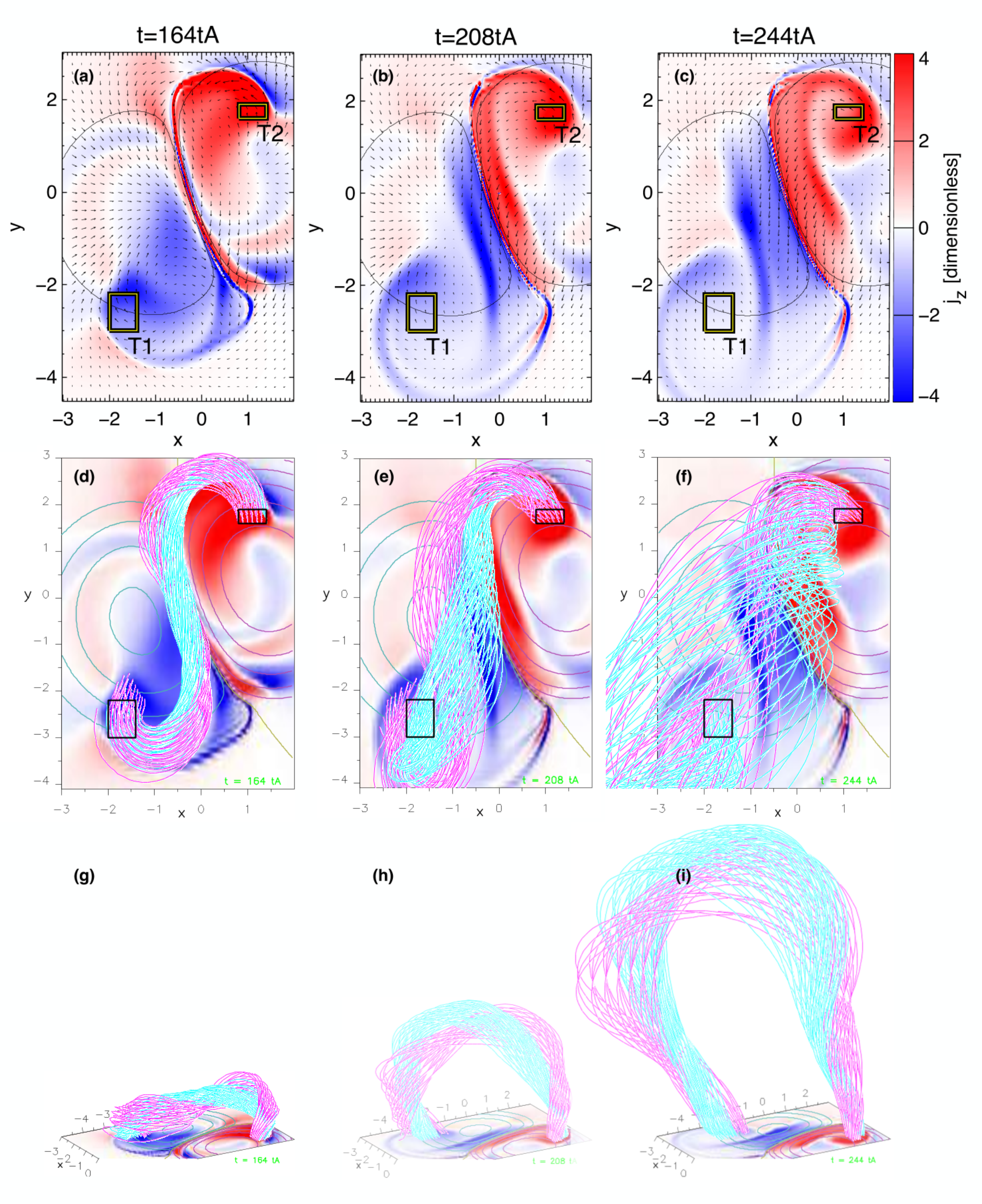}
\caption{The evolution of the simulated vertical component of the photospheric current density $j_z$ (a-i) and the coronal magnetic field structures (d-i). The figure presents the flare simulation directly before magnetic field reconnection ($t=164t_{\mathrm{A}}$), during the eruption ($t=208t_{\mathrm{A}}$), and at a later time ($t=244t_{\mathrm{A}}$). The blue/red color scale presents negative/positive component of the vertical current density $j_{z}$ at $z=0$. The solid-line boxes mark region-of-interests (ROIs T1, T2), which are the flux-rope footpoint-areas. Arrows (a-c) represent the photospheric magnetic field components $B_x$, $B_y$. The black contours (a-c) shows the vertical component of magnetic field ($B_z$) at z = -1 and 1. The magenta and light-blue lines show the magnetic flux rope rooted within the current ribbons' hook presented from top (d-f) and side (g-i) view. An animation of panels (a) through (f) is available. 
The video begins at $t=164t_{\mathrm{A}}$ and ends at $t=244t_{\mathrm{A}}$. The realtime duration of the video is 8 seconds.
\label{fig:f_th}}
\end{figure*}

In Figure~\ref{fig:f_th}, we present the spatio-temporal evolution of the vertical component of the electric current density ($j_z$) at $z=0$ and selected coronal magnetic field structures from a top and a side view. 
The spatial distribution of $j_z$ (Figures~\ref{fig:f_th}a-c) shows J-shaped electric current ribbons in the positive (red) and negative (blue) polarities, where we define as hooks the curved part of the J shape.
In the rest of the map, $j_z$ is significantly lower than in the current ribbons.

Directly before the eruption (Figure~\ref{fig:f_th}a), the straight parts of the opposite polarity current ribbons ($x\in [-1; 1]$ and $y\in [-2.5; 2]$) are parallel to each other and almost symmetric with respect to the polarity inversion line (PIL).
During the eruption (Figure~\ref{fig:f_th}b), the current ribbons move apart from each other and also move away from the PIL.
This trend continues all the way to the end of the simulation (Figure~\ref{fig:f_th}c).
Moreover, the hooks of the electric current density close on themselves with time.
Additionally, during the eruption, the current ribbons become more and more asymmetric.
This asymmetry is related to that of the bipolar field, as well as the expansion of the flux rope, which, instead of expanding upwards, is deflected \citep{2015ApJ...814..126Z}.

\subsection{Flux rope footpoints and flux rope expansion}\label{sec:flux_rope_expansion}

We defined the location of the flux rope footpoints so as to follow the temporal evolution of $j_z$ in these locations.
\citet{2019A&A...621A..72A} showed that the footpoints of the magnetic field lines which belong to the flux rope are encircled by the hook of the electric current ribbon.
Thus, we chose an area located inside the hook of the electric current density ribbons, which is furthermore not disturbed by the evolution of the current density ribbon during the flux rope eruption.
Thereafter, this area is named the flux-rope footpoint-area.
Based on the conditions presented above, we chose two flux-rope footpoint-areas that are presented in two regions of interest T1 (the black-yellow boxes in Figure~\ref{fig:f_th}, $[x;y]\approx[-1.7; -2.6]$) and T2 ($[x;y]\approx[-1.7; -2.6]$ and $[x;y]\approx[1.1; 1.7]$).

Figures~\ref{fig:f_th}d-f present the magnetic field lines of the flux rope indicated by pink/blue lines, which is rooted inside the hooks of the electric current density ribbons.
Initially, the flux rope is composed of an inner core with an arched shape, as shown with the blue lines in Figure~\ref{fig:f_th}g, embedded in an envelope of twisted magnetic field lines as indicated with the pink lines. As the torus-unstable flux rope expands and moves upward, both sets of field lines become stretched (Figure~\ref{fig:f_th}h,i).
A description of the evolution of the magnetic field topology of this simulation is presented by \citet{2015ApJ...814..126Z}, the detailed analysis of the electric current density from $t_1=t_0=164t_A$ to the end of the simulation can be found in \citet{2019ApJ...877...67B}.

In Figures~\ref{fig:f_th}d-i, the pink and cyan contour lines on the photospheric ($z=0$) plane illustrate the positive and negative components of the vertical magnetic field vector ($B_z$). 
When $j_z$ fulfills the condition $j_z/B_z>0$, we refer to this current density as being ``direct'' ($j_z^\text{direct}$).
In contrast, the return electric current density ($j_z^\text{return}$) is used for $j_z$ fulfilling the condition  $j_z/B_z<0$.
The return electric current densities are weaker than the direct one  \citep{2012A&A...543A.110A}.

\subsection{The temporal evolution of electric currents and the electric current density at the flux-rope footpoints}\label{sec:current_evol_th}
While the visual analysis of Figure~\ref{fig:f_th} shows a decrease of $j_z$ in ROI-T1 and ROI-T2 (the ROIs are whiter) with time, we present below a quantitative analysis of this decrease.
For each flux-rope footpoint-area and at each time step, we calculated the average unsigned negative ($j_z^\text{neg}=-{<}{j_z^{<0}}{>}$), positive ($j_z^\text{pos}=<{j_z^{>0}}{>}$) and net ($j_z^\text{net}={<}{j_z}{>}$) electric current density. 
Moreover, we integrated $j_z$ over the entire surface of the flux-rope footpoints area ($S$) to obtain a net electric current ($I_{z}^\text{net}=\iint_{S} j_z^\text{net} ds$).
In the same way, the integration of $j_z^\text{pos}$, ($j_z^\text{neg}$) over the surface where $j_z>0$ ($j_z<0$) allowed us to calculate the positive electric current $I_z^\text{positive}$ (negative $I_z^\text{negative}$).

\begin{figure*}[ht!]
\epsscale{1.0}
\plotone{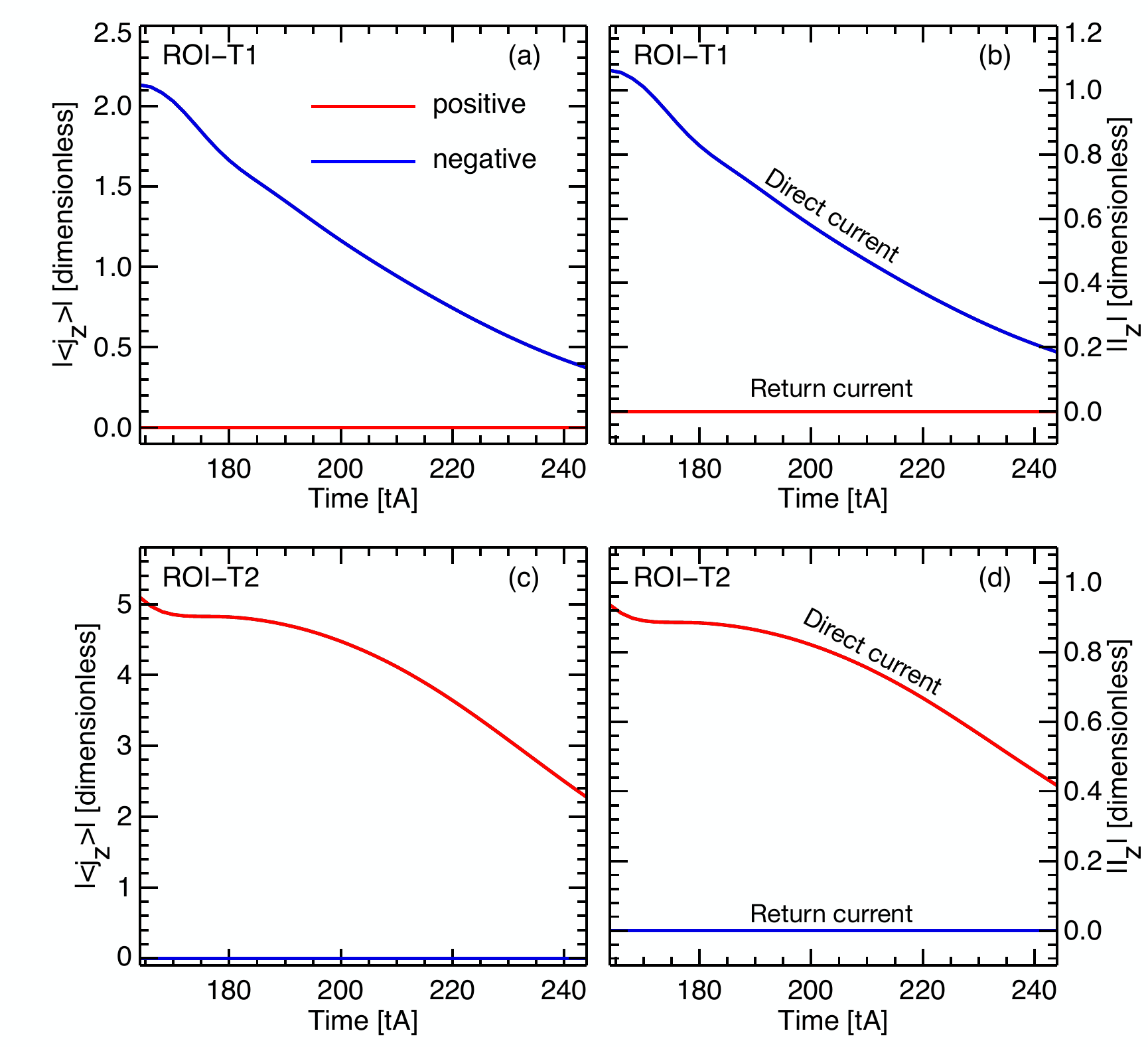}
\caption{Temporal evolution of the vertical component of the electric current density $|<j_z>|$ and electric current $|I_z|$ at the flux-rope footpoint-areas (ROIs T1, T2) is marked in Figure~\ref{fig:f_th}. The top panels (a,b) show $|<j_z>|$ and $|I_z|$ for ROI-T1, the bottom panels (c,d) present the same physical quantity for ROI-T2. The negative and positive components are calculated separately and presented by blue and red lines respectively. 
\label{fig:f_current_th}}
\end{figure*}

In Figure~\ref{fig:f_current_th}, we present the temporal evolution of $j_z$ and $I_z$ at both footpoints areas (ROI-T1 and ROI-T2).
One flux-rope footpoint-area (ROI-T1) is rooted in the strong negative electric current polarity ($j_z^\text{direct}<0$), therefore this dominant current (so-called direct current) is negative ($I_z^\text{direct}<0$).
The positive electric current component at ROI-T1 (so-called return current) is null within the ROI.
The analysis of the temporal evolution of $j_z^\text{direct}$ shows a continuous decrease from the end of the pre-eruptive phase to the end of the simulation (Figure~\ref{fig:f_current_th} a).
We notice the same trend for total direct current (Figure~\ref{fig:f_current_th} b). 
An analogous trend is observed for ROI-T2, however in this case $j_z^\text{direct}$ and $I_z^\text{direct}$ are positive.

The detailed analysis of the temporal evolution of the return current (density) at ROI-T2 shows (Figure~\ref{fig:f_current_th} c,d) a plateau from the beginning of the eruption up to $190t_A$.  
The magnetic field lines rooted in this region are presented in magenta in Figure~\ref{fig:f_th}.
These ``dipped" and ``S-shaped" lines are discussed in more details in Section~\ref{sec:relation_jz_to_L}.

For each ROI, we calculated the absolute (Eq.\,\ref{eq:jz_evolution}) and relative change (Eq.\,\ref{eq:jz_evolution_relative}) of the average $j_z^\text{net}$ and $j_z^\text{direct}$ between the pre-eruptive phase, $t_1=t_0=164t_A$ and the end-of-simulation time $t_2=t_{end}=244t_A$.
\begin{equation}
\Delta j_{z}= j_{z}(t_2)-j_{z}(t_1)\label{eq:jz_evolution}
\end{equation}

\begin{equation}\label{eq:jz_evolution_relative}
\Delta j_{z} (\%)= \frac{j_{z}(t_2)-j_{z}(t_1)}{j_{z}(t_1)}\cdot100\%
\end{equation}
In the same manner, we found the absolute and relative change of $I_z^\text{net}$ and $I_z^\text{direct}$ at the flux-rope footpoint-areas:
\begin{equation}\label{eq:Iz_evolution}
\Delta I_{z}= I_{z}(t_2)-I_{z}(t_1)
\end{equation}

\begin{equation}\label{eq:Iz_evolution_relative}
\Delta I_{z}(\%) = \frac{I_{z}(t_2)-I_{z}(t_1)}{I_{z}(t_1)}\cdot100\%
\end{equation}
In Table~\ref{tab:dd}, we report the results of the $\Delta j_{z}$ and $\Delta I_{z}$ calculations. 
These calculations show a significant decrease of $\Delta j_{z}$ and $\Delta I_{z}$ (more than 80\% for ROI-T1 and more than 50\% for ROI-T2) in the impulsive phase of the flare. 
This strong decrease is clearly visible to the end of the simulation, but we expect that the decrease should continue further up to the end of the flux rope expansion.

\subsection{Relation between the electric current density and the length of the magnetic field line of the flux-tube}\label{sec:relation_jz_to_L}

\begin{figure*}[ht!]
\epsscale{1.0}
\plotone{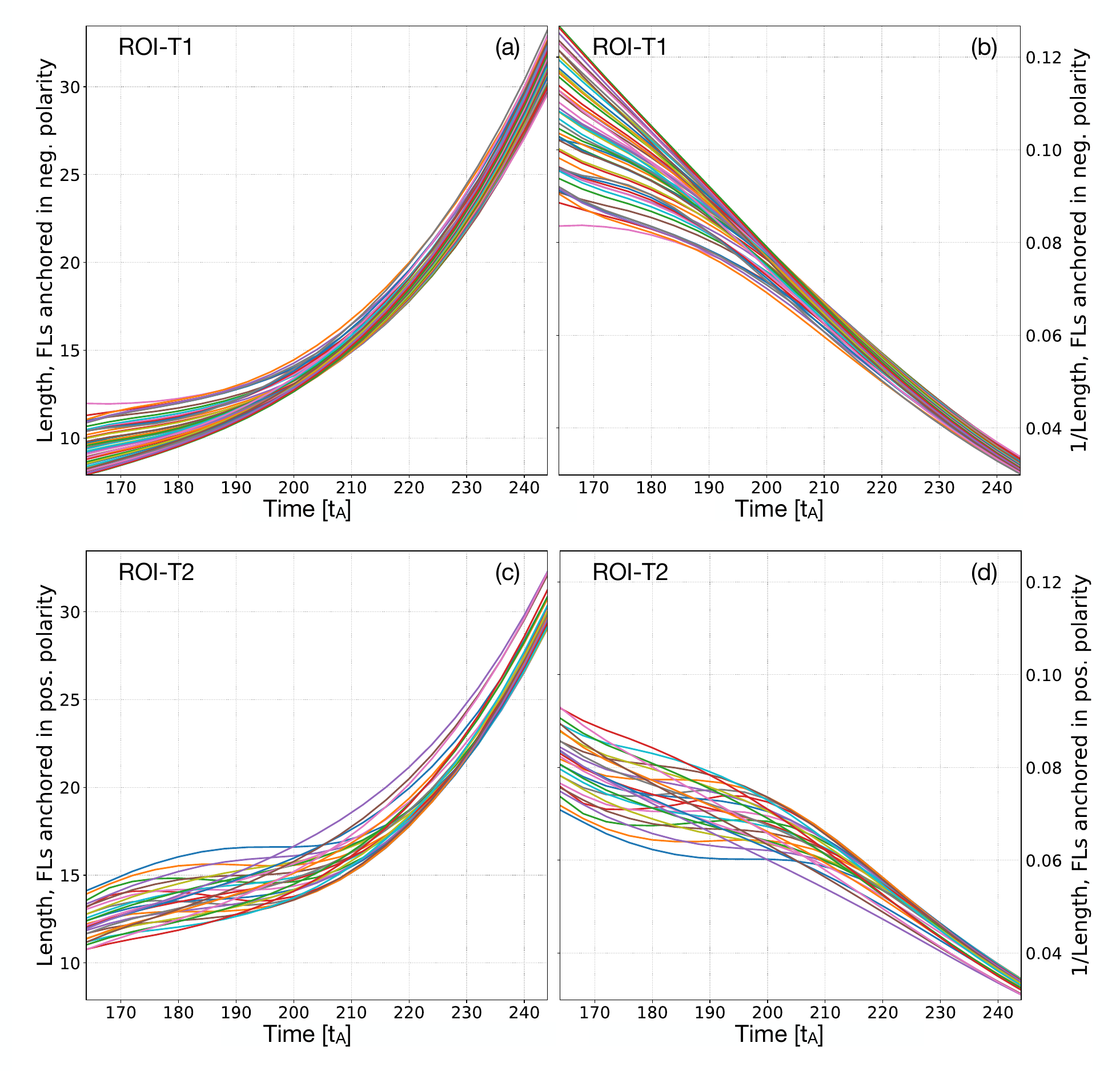}
\caption{The length of the magnetic field line rooted in flux-rope footpoints (ROIs T1, T2) marked in Figure~\ref{fig:f_th}. \nnb{The top panels} (a,b) show the length and reverse length of the magnetic field lines for ROI-T1, the bottom panels (c,d) present the same physical quantity for ROI-T2. These properties are calculated to the magnetic field lines presented in magenta and blue color in Figure~\ref{fig:f_th}}
\label{fig:mag_line_size}
\end{figure*}

In the following, we investigated how the decreasing evolution of the current density at the footpoints of the flux rope is related with the evolution of the flux rope itself. In particular, \nnb{we looked in more detail} at the time evolution of the length of the flux rope field lines. 

In Figure~\ref{fig:mag_line_size}, we show the evolution in time of two different sets of field lines selected from the numerical simulation and rooted at the flux-rope footpoint-areas. These field lines were drawn with the TOPOTR package \citep{Demoulin1996} and correspond to the pink and blue magnetic field lines presented in Figure \ref{fig:f_th} d-i.  In the simulation, which has line-tied conditions at the $z=0$ plane, the same field lines can be tracked in time by fixing one of their two footpoints at the lower boundary. This allowed the study of their morphology and connectivity changes, as shown in \cite{2012A&A...543A.110A, JanvierEtal2013}.

We chose the pink field lines, which represent the core of the flux rope, by fixing their footpoints in the magnetic positive polarity (indicated as the ``T2'' square in Figure\ref{fig:f_th} a-c), while the blue field lines have their fixed footpoints in the negative polarity (``T1'' square in Figure\ref{fig:f_th} a-c). Then, for each selected time step in the numerical simulation, we calculated the $(x,y,z)$ coordinates along each field line in the simulation volume, as provided by the TOPOTR routine. Integrating these coordinates provided the length of each field line, from one footpoint at the $z=0$ plane to another and per time, as reported in Figure~\ref{fig:mag_line_size}.

The evolution of the field line length for the blue field lines is indicated in 
Figure~\ref{fig:mag_line_size}a, while that for the pink field lines (initially the S-shape flux rope field lines) is indicated in Figure~\ref{fig:mag_line_size}c. The inverse of the length is shown in Figure~\ref{fig:mag_line_size}b and Figure~\ref{fig:mag_line_size}d, respectively for the blue and pink field lines, so as to allow direct comparison with the evolution of the current density, as discussed later in Section~\ref{sec:dis_theory}.

These figures show that all field lines increase in length as the simulation goes by, i.e. as the flux rope evolves in time. While this was shown already qualitatively in Figure~\ref{fig:f_th} with the upward stretching of the flux rope between $t=164 t_A$ and $t=244 t_A$, Figure\ref{fig:mag_line_size}a and c show quantitatively how this increase evolves. 

We note that some of the pink field lines (Figure\ref{fig:mag_line_size}c) evolve relatively slowly between  $t=180 t_A$ and  $t=210 t_A$. A look at Figure~\ref{fig:f_th}g-h shows that at $t=164 t_A$, these field lines wind around the blue lines, as opposed to later in time when they become much more stretched upward. This winding of the inner core of the flux rope was already pointed out, in Figure 6 of \cite{2012A&A...543A.110A}, as the typical strongly sheared pre-eruptive field lines often seen as J- or S- shaped sigmoids in observations of active regions. Therefore, while the flux rope core field lines are still evolving, their length does not increase as much during this phase as their morphology changes from highly sheared to vertically stretched (see also Figure~9 in \cite{2012A&A...543A.110A} for a quantification of this change).

Then, from $t=200 t_A$ onward, both sets of blue and pink field lines see their length continuously increase, associated with the vertical stretching of the whole flux rope structure, was shown in Figure\ref{fig:f_th}g-h.

\subsection{Discussion}\label{sec:dis_theory}
For the first time, a 3D MHD simulation is used to study the temporal evolution of the electric current within the flux-rope footpoint-areas. Before our work,  a tentative analysis was performed using observations \citep{ChengDing2016}. The latter authors showed a slight decrease (7-13\%) of the total direct current at the flux-rope footpoint-areas.

Based on the MHD simulation, we found that $j^{direct}_z$, $I^{direct}_z$, $j^{net}_z$, and $I^{net}_z$  significantly decrease (around of 55\% - 82\% compared with the electric current values calculated before the eruption) at the flux-rope footpoint-areas during the impulsive phase of the solar flare (Section~\ref{sec:current_evol_th}).
Therefore, the decrease trend  presented in our simulation is consistent with the  trend of the 
 observations \citep{ChengDing2016}.

The spatio-temporal evolution of the flux rope during its expansion allows us to understand  a possible \nnb{reason for} the  electric currents decrease.
Assuming a simple cylindrical flux tube model (r; z; $\phi$), \citet{AulanierEtal2005A&A...430.1067A} showed that for a cylindrical flux rope of radius \textit{r} and length \textit{L} along the \textit{z}-axis, the electric current density $j_z$ is proportional to the angular component of the magnetic field $B_{\phi}$:\begin{equation}\label{eq:jz_b_phi}
j_z=r^{-1}\partial_{r}(rB_{\phi})\propto B_{\phi}.
\end{equation} 
The magnetic field evolution is described by
\begin{equation}\label{eq:b_l_full}
B_\phi=rL^{-1}B_{z}{\phi}.
\end{equation}
This equation implies that for a constant flux rope radius and a constant end-to-end twist $\phi$, shorter magnetic field lines should be related with a stronger electric current density: 
\begin{equation}\label{eq:jz_l}
j_z \propto L^{-1}.
\end{equation}

We used our simulation to study the relation between $j_z$ and $L^{-1}$ for our expanding flux rope (Section~\ref{sec:relation_jz_to_L}).
%
Figure~\ref{fig:f_current_th}b,d and Figure~\ref{fig:mag_line_size}b,d show that the changes of $j^{direct}_z$ are closely related to changes of $L^{-1}$.
The evolution of $j_z$ at the flux-rope footpoint-areas and $L^{-1}$ are the same, as both significantly decrease during the flux rope expansion. 
The agreement of these trends implies that our simulation is consistent with the previous theoretical analysis \citep[e.g.][]{AulanierEtal2005A&A...430.1067A}.

\nnb{We tested whether the drift of the current ribbons can be responsible for the decrease of the electric current density at the flux rope footpoint area in the ROIs.
The displacement of the surface current ribbons is due to coronal reconnections between the flux rope and surrounding magnetic loops.
If this reconnection for the evolution of currents in the ROIs, then the current variation should be {\it relatively abrupt}.
That can be seen for the same simulation in \citet[Fig5 d]{2019ApJ...877...67B}, where the passage of the ribbon leads to sharp increase followed by sharp decrease of the currents.
Oppositely, if the evolution of the currents are due to the ideal expansion of the flux rope then the current should {\it gradually} decrease (as explain above). 
Our results on the evolution of currents as plotted on Figure~\ref{fig:f_current_th} are consistent with the field line plots of Figure~\ref{fig:f_th}, i.e. the evolution of the electric current in the two ROIs is not related to the surface drift of the current ribbons and flux rope.}

This simulation allows us to answer the other important question of why the electric current (density) measured at the flux-rope footpoint-areas (ROI-T2) shows a plateau from the beginning of the eruption up to 190$t_A$ (Figure~\ref{fig:f_current_th}).
The length \textit{L} of individual flux rope magnetic field lines depends on the distance between their footpoints as well as the flux rope twist.
In the flux rope core, the twist per unit length decreases during the eruption as the flux rope is stretched because the field line expansion is governed by ideal MHD.
However, at the periphery of the flux rope, the twist increases as the result of reconnection.
First of all, the stretching diminishes the twist per unit length in the field lines of the flux rope core, but the total amount of twist should be the same and \textit{L} increases slowly with time (see Figure~\ref{fig:mag_line_size}b,d from $t=164t_A$ to $t=200t_A$), 
When the twist of the flux rope is low, the stretching extends the flux rope length; thus, \textit{L} significantly increases with time (see Figure~\ref{fig:mag_line_size}a,c after $t=200t_A$).
It is worth noting that the pink field lines (see Figure~\ref{fig:mag_line_size}c) are more twisted than the blue lines, thus the plateau effect (Figure~\ref{fig:mag_line_size}b,d) is stronger in ROI-T2 where the pink lines are rooted.

At the beginning of the eruption ($t=165t_A$), the twist and cross-section of the flux rope are larger than at the end of the simulation. 
This implies a large diversity in the  length of the flux rope magnetic field lines at the beginning of the eruption (e.g., Figure~\ref{fig:mag_line_size}a at $t=165t_A$). In other words, some lines are shorter while other lines that are more twisted, are longer. 
Later on, with the stretching all of the field lines and the decrase in the twist of the flux rope core, the lengths of different magnetic field lines become comparable (e.g., Figure~\ref{fig:mag_line_size}a at $t=240t_A$). 

The simulation allows us to understand why the electric current (density) decreases at the flux-rope footpoint-areas, as a result of the flux rope expansion. However, the huge percentage difference of the electric current decrease between the simulation (55\% - 82\%) and previous observations (7\% - 13\%) motivated us to carry out a further observational analysis  (Section~\ref{sec:electric_current_obs_foot}).
These observational analyzes are done for several flaring active regions with the same methods as those used in the simulation.

\section{Observation}\label{sec:obser_and_prep_data}
\subsection{SDO observations}\label{sec:observation}

To study the spatio-temporal evolution of electric current density in active regions, which produced X-class flares, we used data obtained by the Solar Dynamics Observatory \citep[SDO,][]{2012SoPh..275....3P}.
SDO is a space-based mission launched in 2010.
The instruments on-board SDO such as the Helioseismic and Magnetic Imager~\citep[SDO/HMI;][]{2012SoPh..275..207S} and the Atmospheric Imaging Assembly~\citep[SDO/AIA;][]{2012SoPh..275...17L} provide simultaneously magnetic field measurements and full-solar disc images, respectively.

The HMI measures the full Stock parameters that are used to calculate full vector magnetic fields, with a cadence of 12 minutes and spatial resolution of 1 arcsec per pixel (plate scale 0.5 arcsec per pixel).
The vertical component of the electric current density jz is calculated from the HMI vector magnetic field values. 

The AIA is a system of four imaging telescopes that are designed to study the plasma emission from the photosphere to the corona with a high-spatial resolution of 1.2 arcsec per pixel (plate scale 0.6 arcsec per pixel) and a temporal cadence of 12\,s (EUV) and 24\,s (UV).
To identify the footpoints of the flux rope, we used data from all UV and EUV channels, especially in 1600\AA, 193\AA\, and 211\AA.

\subsection{Selection of flare regions}

In our study, we focused on solar flares selected under four specific conditions which reduce considerably the number of possibilities. 

First of all, we needed flare observations obtained with high spectral resolution vector \nnb{magnetograms, so we are limited to} the time range covered by SDO observations (after March 2010). 

Secondly, we needed observations of large sunspots with strong magnetic fields because only these allow us to calculate the electric current density \nnb{from Amp\`ere's law} \citep{SchmiederAulanier2018b} with enough confidence (i.e. above the noise ratio). 
This constraint is because even the best currently used satellite magnetograph has a strongly limited spectral resolution, e.g. HMI provides full solar disk filtergrams at only six different wavelengths \citep{2012SoPh..275..229S}. 
In addition, large sunspots with strong magnetic fields tend to be the source of highly energetic (X-class) solar flares \citep{2000ApJ...540..583S, 2017MNRAS.465...68E,2018SoPh..293...60M}.
This is because the flare energy increases with the sunspot size and the magnetic field strength \citep{2013A&A...549A..66A}.
Therefore we analyzed only X-class flares as they are related to strong magnetic fields.
This very restrictive condition limits our sample since there are only a maximum of 10 to 12 X-class flares observed during the years of intense solar activity \citep{2018SoPh..293...75B}. We considered solar flares observed from 2011 to 2016, leading to 45 X-class solar flares. 

The third condition is to minimize the projection effects. We do so by selecting flares that occur near the disk center (with $x,y\in [-600; 600]$ arcsec), since at these locations the vertical component of the magnetic field is almost parallel to the line-of-sight measured field.
We therefore avoid projection effects that are the strongest close to the solar limb, where the vertical 
component of the magnetic field is almost perpendicular to the line-of-sight.

Based on the XRT Flare Catalog (\url{http://xrt.cfa.harvard.edu/flare_catalog/}), we ended up with a total of 19 flares that satisfied the conditions mentioned above.
These flares are listed in Table~\ref{tab:fp}.

The fourth condition is that we select regions where the flux rope footpoints are easily identified.
In the maps of the electric current density distribution, these regions usually present a J-shaped structure corresponding to the ribbons.
We defined the curved part of the J-shape structure as a hook.
All conditions presented above limited us to nine flares in nine active regions.
%

\subsection{Data preparation}\label{sec:data_preparation}
We used pre-processed SDO data provided by the Joint Science Operations Center (JSOC; \url{http://jsoc.stanford.edu}).
The auxiliary information (e.g.~active region number) was taken from the catalog: Spaceweather HMI Active Region Patch (SHARP; \url{http://jsoc.stanford.edu/doc/data/hmi/sharp/sharp.htm}).

From the JSOC database, we downloaded pre-processed HMI vector magnetograms and AIA images, both with the Lambert Cylindrical Equal-Area projection (CEA), because only the equal-area projection allows us to calculate the average values of the magnetic field and electric current density.
The field-of-view of each observation covers the flaring active region and its surrounding quiet Sun area.
For each flaring region, we exported from the JSOC database four hours of continuous observations - two hours before and two hours after the peak of the X-ray flux related to the flare and observed by the satellite GOES (according to XRT Flare Catalogue).
Thus, the total observation time covers 20 vector magnetic field maps from HMI due to the 12 minute cadence of HMI.

The data obtained from JSOC were corrected \nnb{for solar rotation}.
Four hours of data series of each AIA channels and corresponding HMI vector magnetograms were co-aligned spatially with high-precision as we additionally tested in Section~\ref{sec:sec:data_a_method}. 
Moreover, we checked the mutual co-alignment between data from different AIA channels. 
Finally, we calculated the spatial and temporal alignment between the AIA and HMI data.

\subsection{Co-alignement of the data cubes}\label{sec:sec:data_a_method}
The high-precision alignment of the HMI vector magnetograms and AIA data from all channels is critical in our study because it has a direct influence on the identification of the flux-rope footpoint-areas.
Therefore, we tested the co-alignment precision of the pre-processed data from the JSOC database.

To this aim, in each ROI, we chose the quiet Sun region, rich in network structures; then, using a cross-correlation method we co-aligned together all AIA and HMI data.
We kept the original resolution of HMI data because we need this data to calculate the spatial derivative of $\bm{B}$ to obtain ($\bm{\nabla} \bm{\times} \bm{B}$) with the highest possible accuracy.
AIA images  are mainly used   for the identification of the flux rope footpoints. Therefore, we interpolated AIA data (plate scale 0.6 arcsec per pixel) to the HMI plate scale (0.5 arcsec per pixel).
Finally, the maps of all components of the vector magnetograms ($B_{x}$, $B_{y}$, $B_{z}$) were shifted, but only by an integer number of pixels, the maximal shift was $\pm2$ pixels in the x-direction and y-direction."

The electric current density ($j_z$) was 
derived from
%
the horizontal components of the HMI vector magnetic field ($B_x$ and $B_y$),  
applying 
the centered difference method with \nnb{Amp\`ere's law}~(\ref{eq:ampere_law}), such as in the simulation. 
To calculate the electric current $I_z$, we computed the area of the region of interest (ROI) in physical units (at the disc center, 0.03 degree per pixel=0.5 arcsec per pixel=363 Mm per pixel according to \citet{2013arXiv1309.2392S}).

In summary, after applying the procedure presented above, we obtained a spatio-temporally co-aligned sequence of 4 hours of data of all components of HMI vector magnetograms ($B_x$, $B_y$, and $B_z$) and all AIA channels, for 19 flares.

\subsection{Flux-rope footpoints identification -imaging data analysis}\label{sec:footpoint_intentificatio_methods}

We investigated the temporal evolution of the electric current density in erupting flux-rope footpoint-areas after defining the ROIs for the  pre-selected 19 flares using HMI and AIA data cubes (Table~\ref{tab:fp}).

Our simulation and previous studies \citep{2014ApJ...788...60J, 2014ApJ...784..144D, 2019A&A...621A..72A} showed that flux-rope footpoint-areas are surrounded by the hooks of the electric current ribbon (footprints of the associated quasi-separatrix layers (QSLs)).
Thus, we looked for the hooks of the electric current ribbons to find the associated flux-rope footpoint-areas.
\nnb{To the identification of the flux flux-rope footpoint-area we used the electric current density maps together with corresponding intensity maps and we based on the statistical methods (visual cross-correlation of the $j_z$ trends between neighbouring pixels within the ROIs and check that $j_z$ variation is larger than noise, see Sect.~\ref{sec:average_jz}).}
In some active regions, the identification of these hooks is almost impossible, because the magnetic field topology is too complex, or the noise level is too high, or the electric current ribbon expands too fast over a large surface. 
Therefore, we also searched for the flux-rope footpoint-areas using the AIA  data cubes.

\citet{2014ApJ...788...60J, 2016A&A...591A.141J} and \citet{Savcheva2016} showed that the shape of the electric current ribbons corresponds well to the shape of the photospheric ribbons, when observed in AIA 1600\AA~images.
Thus, we analyzed the 1600\AA\ images to find the hooks of photospheric ribbons.
However, in some active regions, the photospheric ribbons have a complex shape (e.g.  with many bends).
In those cases, the identification of hooks can also be difficult or even impossible.
Therefore, we used AIA 1600\AA~images together with coronal images (e.g. AIA 211\AA) to look for other structures or phenomena that can be useful to identify the flux rope position.
%

For example, coronal images of flaring regions usually show transient coronal holes, also named coronal core dimmings, which are often associated with flux-rope footpoint-areas \citep{1997ApJ...491L..55S, 1998GeoRL..25.2465T}.
In those regions, the intensity decreases in EUV and SXR as well as the density due to the loop expansion during the CME outward motion.
However, the coronal dimming regions do not only develop around the flaring PIL, but also in different other areas \citep{2001JGR...10629239K, 2019A&A...621A..72A}, therefore the dimming regions can not be independently used to determine the position of the flux-rope footpoint-areas.
In order to identify the flux-rope footpoint-areas, we took into account the location of the coronal core dimming as well as  the shape of the electric current ribbons and the UV hook ribbons.

\subsection{Flux-rope footpoints identification -average $j_z$ and $I_z$ in the flux-rope footpoint-areas}\label{sec:average_jz}

In some active regions, the complex ribbon topology and wide coronal core dimming hamper an unambiguous identification of the flux-rope footpoint-areas.
For these ROIs, we used an additional method based on the average value of electric current (density) and its maximal error, both calculated in the flux-rope footpoint-areas surrounding.
This method was tested in ROIs where the flux-rope footpoint-areas are clearly defined in the electric current density map (e.g. ROI-1A and ROI-1B in Sect.~\ref{sec:AR11158}).

First, we defined the approximate localization of the flux-rope footpoint-areas using hooks of the electric current ribbons, the photospheric ribbon hooks, and the coronal core dimming position, such as in Sect.~\ref{sec:footpoint_intentificatio_methods}.
The surface around this area was divided into 16 equal-size squared sub-regions, where the size of individual sub-regions was comparable or slightly smaller than the ribbon hooks diameter.
In each sub-region, we computed the average direct, return and net $j_z$ and $I_z$.
These computations were provided with a 12 min time step and covered two hours before and two hours after the peak of the X-ray flare emission.
Additionally, we calculated the $j_z$ error for each time step:
\begin{equation}\label{eq:error_general}
\delta (j_{z})= \frac{3\sigma}{\sqrt{n}},
\end{equation}
where $\sigma$ is the  standard deviation of $j_z$ inside each grid cell at a defined time-step; $n$ is the area of the single grid in pixels.  
For each grid cell, we calculated the absolute and relative changes of the electric current density ($\Delta j_z$) comparing $j_{z}(t_{1})$ in the pre-eruptive phase with $j_{z}(t_{2})$ in the post-eruptive phase (such as in Sect.~\ref{sec:current_evol_th}) using Equation\ref{eq:jz_evolution} and  Equation\ref{eq:jz_evolution_relative}, respectively.
Moreover, we calculated the maximal error of ($\Delta j_z$):
\begin{equation}\label{eq:error_max_jz}
\sigma_{max}(j_z)= |\sigma (j_z, t_1)|+|\sigma (j_z, t_2)|.
\end{equation}
In the same manner, we computed the relative change of $I_z$.
The error of $I_z$ was calculated as:
\begin{equation}\label{eq:error_max_Iz}
\sigma_{max}(I_z)= |\sigma (I_z^{t_1})|\sigma_S+|S|\sigma (j_z^{t1})+|\sigma (I_z^{t_2})|\sigma_S+|S|\sigma (j_z^{t_2}).
\end{equation}
The temporal evolution of $j_z$ and $I_z$  shows a clear trend, but only in a single sub-region (1 of 16) of some ROIs; in the other sub-regions or even all sub-regions of some ROIs, the changes are significantly smaller than the maximal error.

In the next step, we identified the place with the highest $j_z$ and $I_z$ change.
To this aim, we shifted a whole grid with an offset of the half-size of a single grid (half hook size) in x- and y-direction to check that in the new position, the changes of the electric current density are more/less evident.
The center position of the sub-region with the highest $j_z$ and $I_z$ evolution and the smallest maximal error was selected as the center of the new circular sub-region.

We used a circular sub-region because the simulation and observations show that the (electric) ribbon hooks are circular.  
To find an optimal radius size and the position of the circular sub-region, we modified the radius and shifted the position of the circle center with x and y directions with a distance smaller than the ribbon hook size.
For each modification, the average $j_z$ and $I_z$ and their maximal errors were calculated.
It is important to note that on the one side, the increase of the sub-region radius reduces the maximal error, such as implied by Equation~\ref{eq:error_general}.
On the other side, when the radius is too large the influence from the noisy region out of the flux-rope footpoint-areas strongly increases the maximal error.
The increase of the maximal error caused by the counting signal out of the flux-rope footpoint-areas is significantly stronger than the maximal error decrease caused by increasing the sub-region radius.
Thus, we assumed that the circular sub-region covers the flux-rope footpoint-areas when the change of $j_z$ (or $I_z$) is the highest, and the maximal error is the lowest.

The method mentioned above was used to calculate the evolution of the electric current (density) for the flux-rope footpoint-areas in Section~\ref{sec:electric_current_obs_foot} and in Appendix~\ref{sec:appendix}.
However, in several ROIs the identification of the flux-rope footpoint-areas was difficult.

\section{Electric current (density) evolution at the flux-rope footpoints}\label{sec:electric_current_obs_foot}

We discuss in the following the evolution of the electric current (density) for three different flares that are the most representative.
For our analysis, we used the method described in Sections~\ref{sec:footpoint_intentificatio_methods} and \ref{sec:average_jz} to define the ROI of each flare. 

\subsection{AR 11158}\label{sec:AR11158}
We start the analysis with the active region NOAA AR 11158 because this active region has a relatively simple magnetic field configuration with a bipole in the central part
\citep{2011ApJ...738..167S}
as in our 3D MHD simulation (see Section~\ref{sec:footpoint_intentificatio_methods}). Besides, the ribbons of one main flare  occurring in this  region on 15th February 2011 have  been well analysed  by us   considering the UV emissions as well as the electric current density \citep{2014ApJ...788...60J}. This region was the starting point of our research  concerning the identification of  flux rope and the  footpoint locations due to  its clear magnetic topology.
Between the 13th and 21st February 2011, this active region was a source of 63 flares.
The flare occurring on February 15 was one of the strongest, of class X2.2.
On that day the AR 11158 had a quadruple magnetic field topology (outer and central bipoles) and was located nearby the solar disk center \citep{2012ApJ...748...77S}.
Several authors studied this flare in depth, e.g.~\citet{2012ApJ...748...77S, 2013SoPh..287..415P, 2012ApJ...749...85G, 2012ApJ...745L..17W, 2013ApJ...770....4T, 2014ApJ...788...60J, ZhaoEtal2016, 2015A&A...580A.106M, 2015ApJ...811...16K}.

\begin{figure*}[ht!]
\epsscale{1.0}
\plotone{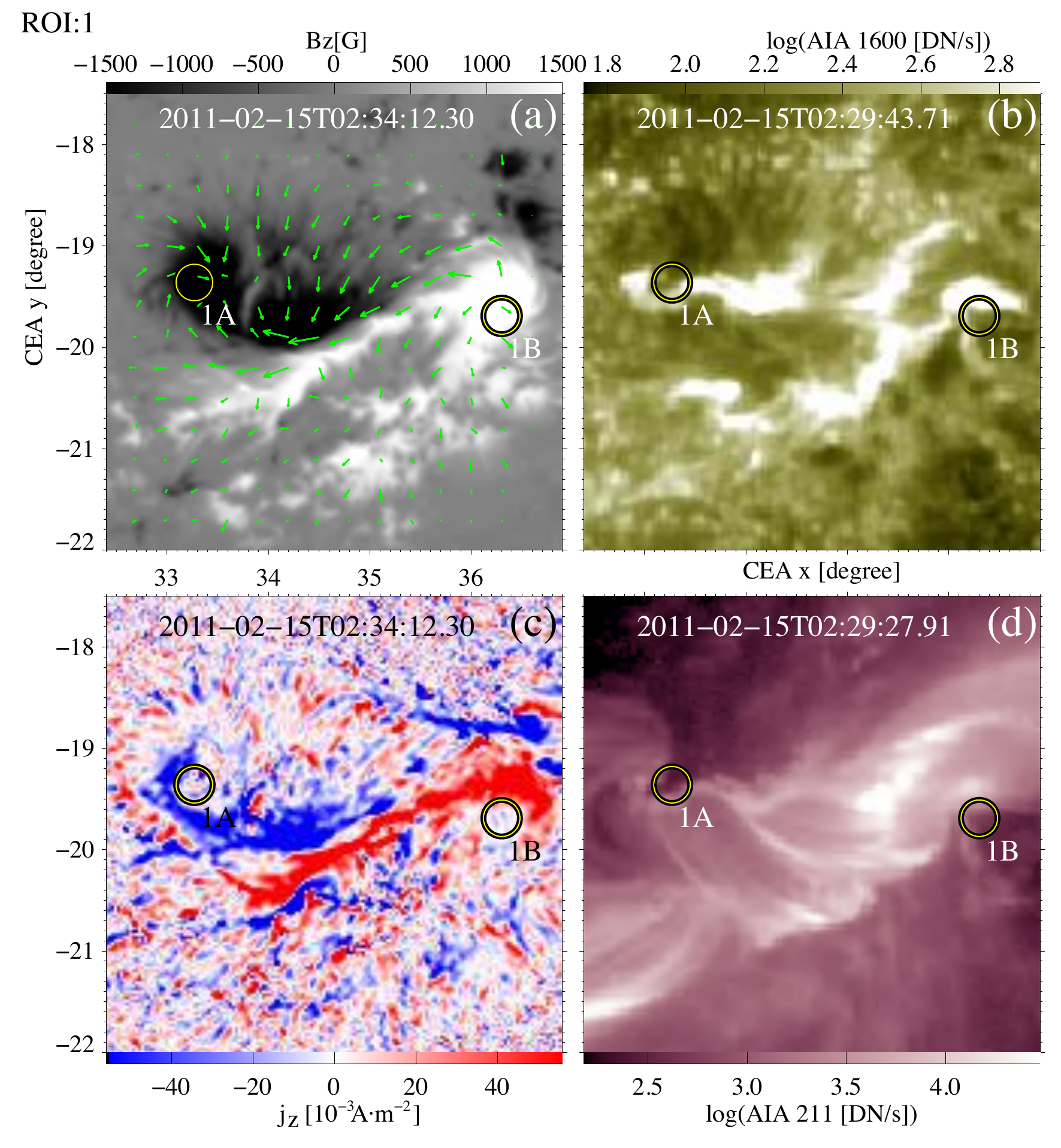}
\caption{Context SDO images for the X2.2 class flare event area of 15th February 2011 (after the eruption); (a) HMI vector magnetogram, (b) AIA 1600\AA, (c) the vertical current density map, (d) AIA 211\AA. The footpoints of the flux rope are marked by a yellow-black circle (ROIs 1A, 1B). Arrows (a) represent the horizontal photospheric magnetic field components $B_x$, $B_y$, the vertical component of the magnetic field is defined by the white/black color scale.
An animation of this figure is available. The video begins on Feburary 14, 2011 at approximately 23:59:00 and ends the next day around 03:59:00. The
 realtime duration of the video is 48 seconds. \label{fig:roi1f}}
\end{figure*}

Figure~\ref{fig:roi1f} and the associated movie (Movie1) summarize the observations of this  flare  as well as the evolution of the AR during four hours.
Figure~\ref{fig:roi1f} (a) shows the vertical magnetic field  of the  central bipole clearly concentrated into two compact structures of opposite polarities ($[x;y]\approx[-19^{\circ}; 33^{\circ}]$ and $[x; y]\approx[-19^{\circ}; 36^{\circ}]$).
During the flare, the vertical  magnetic field  presents only minor changes, as shown in the movie while the horizontal magnetic field increases between the two dominant magnetic field polarities (see \citet{2017ApJ...839...67S}).

In the same region of interest, the AIA 1600\AA~ channel~(Figure~\ref{fig:roi1f} b) clearly shows two bright and elongated flare ribbons.
Their central parts are parallel, and each ribbon ends with a hook.
The associated movie shows that the ribbons spread away from each other during the eruptive phase, and the hooks strongly change shape.

The location, shape, and evolution of the flare ribbons (Figure~\ref{fig:roi1f} b) are similar to those of the electric current ribbons (Figure~\ref{fig:roi1f} (c) and  associated movie).  This figure is consistent with the figure 6 of \citet{2014ApJ...788...60J} representing the J-shape flare ribbons and the electric current  density ribbons which they interpreted by the presence of a flux rope and quasi-separatrix layers (see their scheme in their  figure 7). In their paper, the authors explained the increase of the electric current density  inside the ribbons  by  the collapse of the  flare's current sheet layer  which is  an independent structure of the flux rope.
In the two opposite polarities, J-shaped current ribbons (red-positive and blue-negative) also ended with a clear hook each (Figure~\ref{fig:roi1f} (c)).
The straight part of the current ribbons is almost symmetric with respect to the PIL. 
The electric currents ribbons are compact structures with $j_z$  significantly higher than the noise level, and they exist during the whole analysis time.
%

In this active region, the electric current density map can be easily used to identify hooks and the flux-rope footpoint-areas inside the hooks (see the yellow-black circle) using the method described in Section~\ref{sec:footpoint_intentificatio_methods}.
However, we used this clear example to also show that hooks of the flare ribbon of AIA 1600\AA~channel can be used to identify the flux-rope footpoint-areas.
Looking at the other AIA channels, we found two clear coronal core dimming regions in AIA 211\AA~ (Figure~\ref{fig:roi1f} d) at the flux-rope footpoint-areas.
However, the AIA 211\AA~ can only be used to identify the flux-rope footpoint-areas together with AIA 1600\AA~ as discussed in Section~\ref{sec:footpoint_intentificatio_methods}.

\begin{figure*}[ht!]
\epsscale{1.0}
\plotone{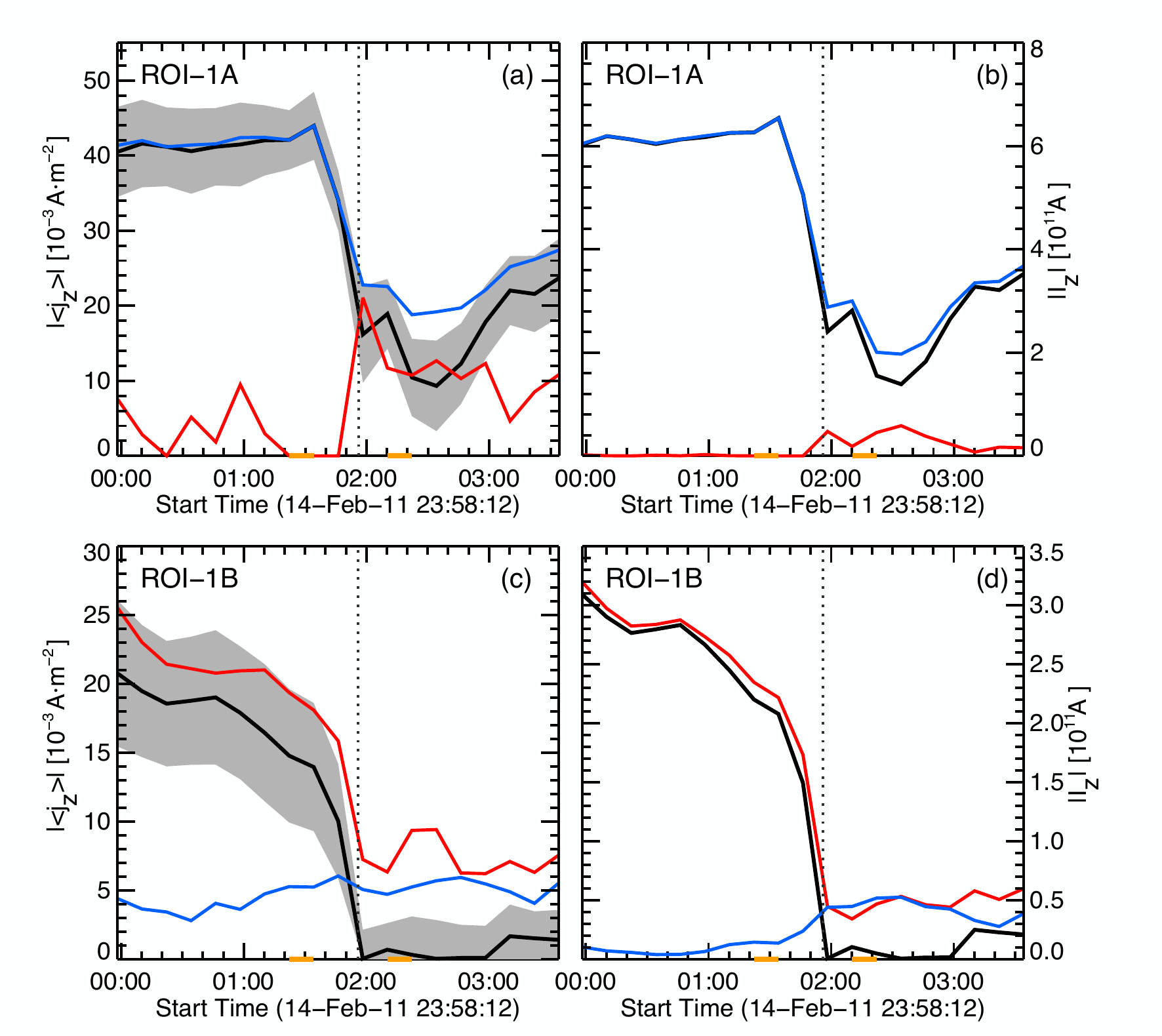}
\caption{
The temporal evolution of the vertical component of the electric current density $|<j_z>|$ and electric current $|I_z|$ at the footpoints of the flux rope (ROIs 1, 2) marked in Figure~\ref{fig:roi1f}. \nnb{The top panels} (a,b) show $|<j_z>|$ and $|I_z|$ for ROI-1, the bottom panels (c,d) present the same physical quantity for ROI-2. The solid line presents negative (blue), positive (red) and net (black) component of $|<j_z>|$(a,c) and $|I_z|$(b,d). 
The grey field indicates the range of the maximal error of the net  $|<j_z>|$(a,c). The vertical dotted line indicates the peak of the flare. The orange marks indicate the time steps used to calculate the decrease of $|<j_z>|$(a,c) and $|I_z|$ (see Table~\ref{tab:dd}).
 \label{fig:roi1c}}
\end{figure*}

Based on the same methods as used in Section~\ref{sec:current_evol_th} with the MHD model, we calculated the average electric current density and electric current for each flux-rope footpoint-areas (ROI-1A and ROI-1B) at each timestep. 
In Figure~\ref{fig:roi1c}, we present the temporal evolution of the direct, return, and net components of $j_z$ and $I_z$. 
For ROI-1A and ROI-1B, we found that the absolute value of the net and direct electric current density decrease during the flare eruption, but the return current slightly increases (ROI-1A) or stays almost constant (ROI-1B).
In both cases, the direct current decreases sharply during the impulsive phase. 
Comparing the pre-eruptive phase and time after the impulsive phase, we calculated (Equation~\ref{eq:jz_evolution_relative}) that the relative decrease of the net and direct electric current density are larger than 50\% (Table~\ref{tab:dd}) in both flux-rope footpoint-areas.
This decrease is significantly higher than the maximal error (see the grey area in Figure~\ref{fig:roi1c}).
The same trends, such as described above, are observed for the electric currents (Figure~\ref{fig:roi1c}, b, d).
However, the changes of the electric current at the flux-rope footpoint-areas are different (trend and magnitude) than the $I_z$ increase inside the electric current ribbons analyzed by \citet{2014ApJ...788...60J}. 

\subsection{AR 12205}\label{sec:AR12205}
The second example is the active region NOAA AR 12205 with a complex magnetic field topology and a flare with a weak magnetic field inside the ribbon hooks. 
In this region, the identification of the flux-rope footpoint-areas is only possible based on the imaging data from SDO/AIA (see Section~\ref{sec:footpoint_intentificatio_methods}).
This region produced 56 flares between 3rd and 17th November 2014, but only one X-class flare.
%
We analyzed the X1.6 class flare that was observed between 16:53UT and 17:34UT on the 7th November 2014.
This flare was studied in depth by \citet{2015ApJ...812..172Y} and \citet{2017ApJ...839...67S}.

\begin{figure*}[ht!]
\epsscale{0.84}
\plotone{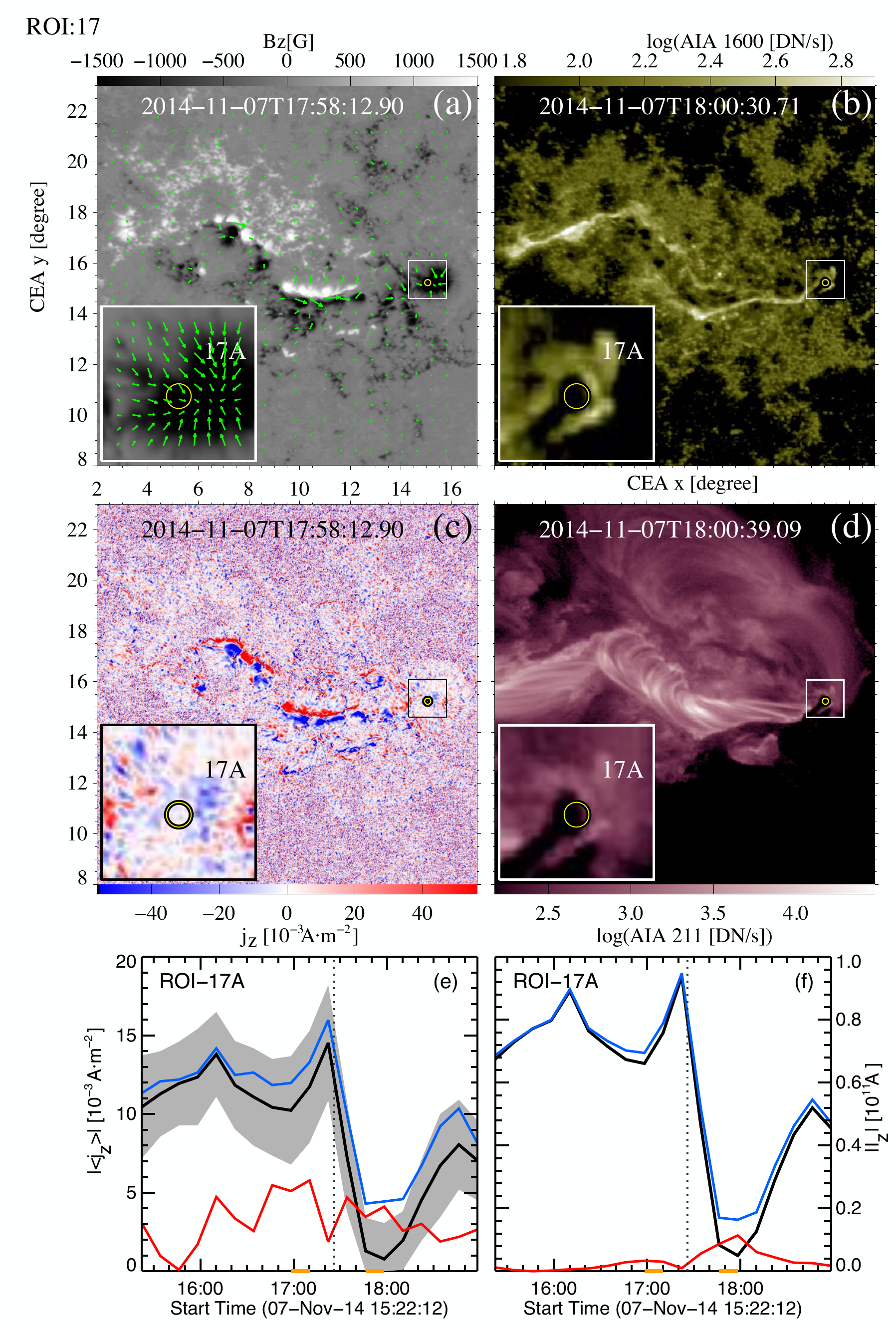}
\caption{Context SDO images for the X1.6 class flare event area of 7 November 2014 (after the eruption) and temporal evolution of the electric current and electric current density at the flux-rope footpoint-areas; (a) HMI vector magnetogram, (b) AIA 1600\AA, (c) the vertical current density map, (d) AIA 211\AA, (e) temporal evolution of the electric current density at the flux-rope footpoint-areas, (f) temporal evolution of the electric current at the flux-rope footpoint-areas.
The insets in the left-bottom corner shows zoomed-in selected area (smaller box) around the footpoints of the flux rope. The footpoints of the flux rope are marked by a yellow-black circle (ROI 17A). 
The photospheric magnetic field map (a) is presented in the same manner as in Figure~\ref{fig:roi1f}. Plots of the temporal evolution of the electric current density (e) and electric current (f) are presented in the same way as plots (a,c) and (b,d) in Figure~\ref{fig:roi1c}, respectively.
An animation of panels (a) - (d) is available. The video begins on November 7, 2014 at approximately 15:23:30 and ends the same day around 19:23:30. The
 realtime duration of the video is 48 seconds.
 \label{fig:roi17f}}
\end{figure*}

The complex magnetic field configuration, as presented on Figure~\ref{fig:roi17f} (a), implies a more complex shape of the flare ribbons (Figure~\ref{fig:roi17f}b) than those in AR 11158. 
Figure~\ref{fig:roi17f}(b) presents two well-separated flare ribbons observed in AIA 1600\AA, and each of them ends with a hook.
The hook of the north-east ribbon (Figure~\ref{fig:roi17f}b) and the corresponding electric ribbon hook (Figure~\ref{fig:roi17f}c) are poorly developed; therefore, it was difficult to define the flux-rope footpoint-areas. 
The south-west flare ribbon ends with a long hook (see the zoomed area in Figure~\ref{fig:roi17f}b), but the magnetic field is weak there; hence it implies strong noise of the electric current density (see the zoomed area in Figure~\ref{fig:roi17f}c).
Therefore, we used the clear location of the flare ribbon hooks in AIA 1600\AA~ map (Figure~\ref{fig:roi17f}c) and auxiliary the position of the coronal core dimming region from AIA 211\AA~(Figure~\ref{fig:roi17f}d) to define the flux-rope footpoint-areas (yellow-black circle) with a method described in Sections~\ref{sec:footpoint_intentificatio_methods} and \ref{sec:average_jz}.

At the flux-rope footpoint-areas, the statistical study shows a clear drop of the direct (in blue) and net (in black) electric current density during the flare (Figure~\ref{fig:roi17f}f).
Despite a large noise in an individual pixel of the electric current density map, the decrease of the $j_z$ and $I_z$ is higher than the maximal error (gray error bars).
The same trend presents $I_z$.
In ROI-17A, $j_z^\text{direct}$ and $I_z^\text{direct}$ dropped by 65\% and 77\%, respectively. 
Moreover, $j_z^\text{net}$ and $I_z^\text{net}$ decreased by 91\% (Table~\ref{tab:dd}).
However, during the decaying phase, we noticed that after a clear minimum of around 18:00 UT, the net and direct $j_z$ and $I_z$ slightly increase. 
In contrast, the return $j_z$ and $I_z$ (in red) stay almost constant before, after, and during the flare.

\subsection{AR 12297}\label{sec:AR12297}
As the third example, we present the active region NOAA AR 12297 with a complex current density map and a quickly evolving hook that make impossible a  direct identification of the flux-rope footpoint-areas without  SDO/AIA observations.
This active region  was the source of 138 solar flares from the 5th to  the 21st March 2015.
We focused on the X2.1 class flare, the strongest one generated by this region, observed on 11th March 2015.
This flare was deeply studied by \citet{2017ApJ...839...67S}, \citet{2019ApJ...876..133L} and \citet{2016ApJ...830..152L}.

\begin{figure*}[ht!]
\epsscale{0.84}
\plotone{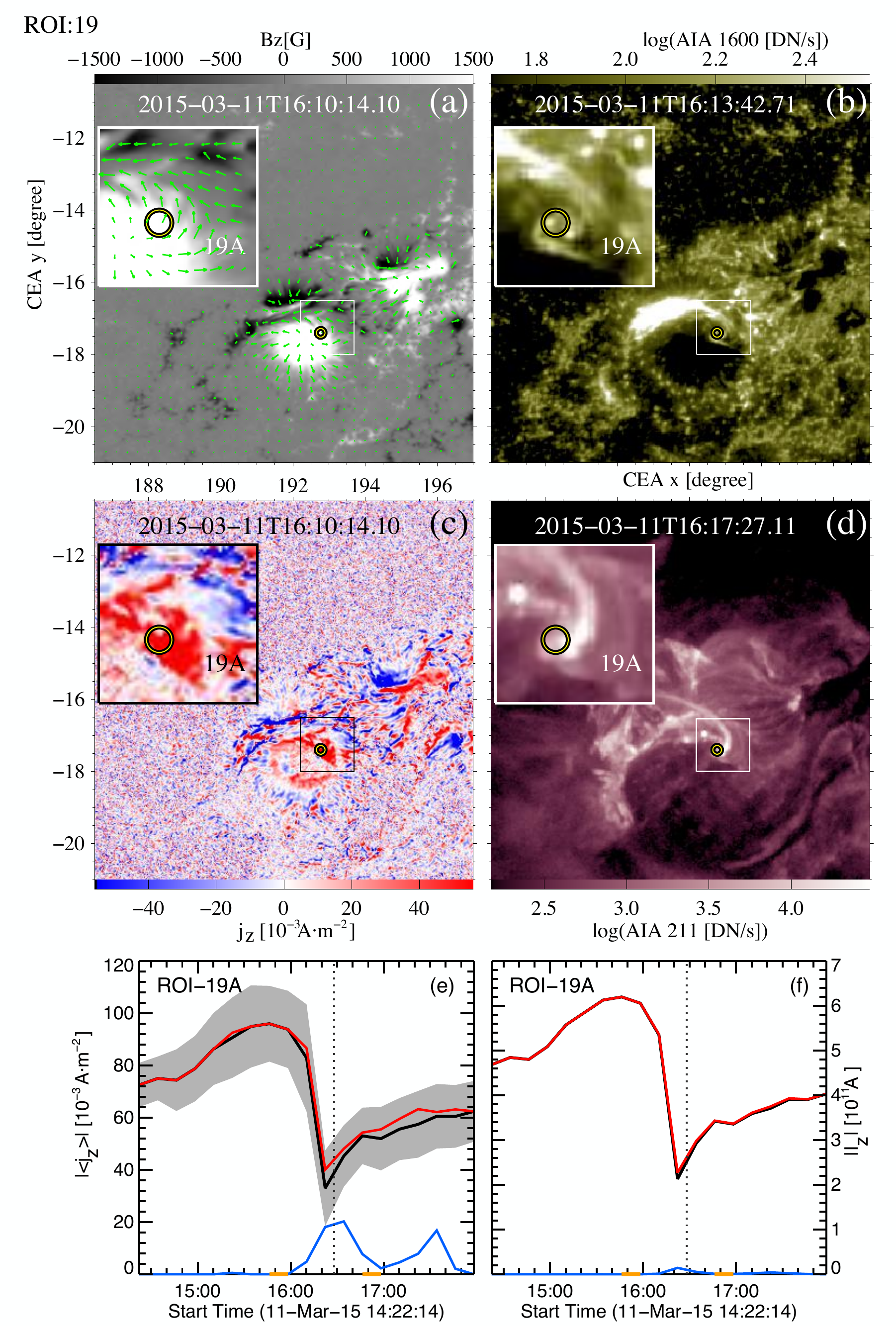}
\caption{Context SDO images for the X2.1 class flare event area of  11 March 2015 and temporal evolution of the electric current and electric current density at the flux-rope footpoint-areas; (a) HMI vector magnetogram, (b) AIA 1600\AA, (c) the vertical current density map, (d) AIA 211\AA, (e) temporal evolution of the electric current density at the footpoint of flux rope, (f) temporal evolution of the electric current at the footpoint of flux rope.
The insets in the left-bottom corner shows zoomed-in selected area (smaller box) around the footpoints of the flux rope. The footpoints of the flux rope are marked by a yellow-black circle (ROI 19A). 
The photospheric magnetic field map (a) is presented in the same manner as in Figure~\ref{fig:roi1f}. Plots of the temporal evolution of the electric current density (e) and electric current (f) are presented in the same way as plots (a,c) and (b,d) in Figure~\ref{fig:roi1c}, respectively.
An animation of panels (a) - (d) is available. The video begins on March 11, 2015 at around 14:24:30 and ends the same day at approximately 18:28:00. The
 realtime duration of the video is 48 seconds.
\label{fig:roi19f}}
\end{figure*}

Figure~\ref{fig:roi19f}a presents the  quadrupolar magnetic field configuration \nnb{of the AR}.
The flare occurs nearby the south-east dipole, which is strongly asymmetric.

Using AIA~1600\AA\,(Figure~\ref{fig:roi19f}b) data, we identified two ribbons.
The southern ribbon is ended with a quickly evolving hook; the northern one has a complex shape that does not allow \nnb{one to determine a hook}.

The electric current density map (Figure~\ref{fig:roi19f}c) presents a system of two pairs of opposite polarity electric current ribbons.
The first pair is located in the northeast part of the active region and stays almost the same during the whole observation time.
The second pair is in the southwest part of the active region.
The opposite polarity current ribbons move away from each other during and after the impulsive phase.
This effect was also noticed by \citet{Savcheva2016} and in ROI-17 and ROI-19 of our paper.
This movement was previously noticed in \citet{Savcheva2016}.
In this active region, the direct identification of the hook of the electric  current ribbon is impossible because of the numerous small-scale structures concentration of the electric current. 
Therefore, we used the AIA 1600\AA~ map (Figure~\ref{fig:roi19f}b), the S-shaped loops from AIA 211\AA, and the location of the coronal core dimming (see AIA 211\AA~ movie) as the proxy for ribbon hooks.

The analysis of the temporal evolution of the current (density) shows the direct $j_z$ (Figure~\ref{fig:roi19f}e) and $I_z$ (Figure~\ref{fig:roi19f}f) decrease sharply (40\% and 45\%, respectively) around 10 minutes before the flare peak;  the net $j_z$ and $I_z$ present the same trend.
The return $j_z$ and $I_z$ stay almost constant.

\subsection{Other regions of interest}\label{sec:otherar}
We studied 19 ROIs (Table~\ref{tab:fp}) in total; three of the most representative ROIs are described already in Section~\ref{sec:AR11158}-\ref{sec:AR12297}.
We analyzed the other 16 ROIs using the same methods presented in Sections~\ref{sec:footpoint_intentificatio_methods} -  \ref{sec:average_jz} in order to make statistics on the trend of the electric current in the hooks. Each of the 16 ROIs presents a magnetic field topology, electric current density configuration, ribbon, and coronal core dimming localization that are  similar to the examples presented in Section~\ref{sec:AR11158}-\ref{sec:AR12297}.

However for ten of them \nnb{the hook's identification} was difficult because of their complex magnetic field topology and the weak magnetic field in the hook surrounding that implies a high noise level in the electric current density maps.
In these ten ROIs,
we 
often 
found a quickly drifting (or drift) and highly curved flare ribbons, such as the southern ribbon in AR 12205 (Section~\ref{sec:AR12205}), or was difficult to find a ribbon at all, such as the second ribbon like in AR 12297 (Section~\ref{sec:AR12297}).  

For three others ROIs, we identified only one single flux-rope footpoint-areas.
For this identification, we compared the position of the ribbon hooks (AIA 1600\AA) and the coronal core dimmings (AIA 211\AA).
Despite the electric current (density) decreases at the flux-rope footpoint-areas for  these three ROIs.
However the changes are slightly smaller than the maximal error of $j_z$ and $I_z$; hence they are not considered in the further analysis.

In the last three ROIs, we identified the four flux-rope footpoint-areas; in three of them, the electric current (density) shows a clear decrease during a flare.
These three events are deeply discussed in Appendix~\ref{sec:appendix}.
All this  precise analysis  of the nine  flaring active regions allows us to make the following statistical analysis (see Table~\ref{tab:fp}).

\subsection{Statistical analysis of the electric current density decrease at the flux-rope footpoint-areas}\label{sec:trend}
We analyzed 19 ROIs, and
found 11 clear ribbon hooks after a long search of identification using all our techniques. For seven of them we noticed a clear decrease of the electric current (density) inside the hook, $j_z$ and $I_z$,  larger than the maximal error during the impulsive phases (Table~\ref{tab:dd}).
In the other four  ROIs
the decrease was smaller than the maximal error and we put 0 in Table~\ref{tab:dd}.

Despite the small sample that we have,  a statistical study can be done.
For all the investigated flux-rope footpoint-areas the median of $j_z^\text{direct}$ and $I_z^\text{direct}$ decrease is 52\% and 65.1\%, respectively; 
for $j_z^\text{net}$ and $I_z^\text{net}$ it is 61\% and 61\%, respectively. 
The median decrease of $j_z^\text{direct}$ and $I_z^\text{direct}$ calculated in all  flux-rope footpoint-areas  is 52\% and 65.1\%, respectively;  for $j_z^\text{net}$ and $I_z^\text{net}$ it is 61\% and 61\%, respectively. 
On the contrary, the return current presents only minor changes or stays constant during the whole flare eruption time.

\setcounter{table}{0} 
\begin{table*}[h!]
\renewcommand{\thetable}{\arabic{table}}
\centering
\caption{Properties of flares, ribbon and ribbon hooks visibility for 19 flares. The two last columns present the numbers of the identified hooks (Hooks clear) and the numbers of hooks where a decrease of the electric current (density) during the eruptive phase is larger than the maximal error (Hook with $j_z$ decrease).} \label{tab:fp}
\begin{tabular}{lllllcc}
\tablewidth{0pt}
\hline
\hline
ROI & Event Peak Time [UT] & AR & Magnitude & Flare type   & Hooks clear & Hook with $j_z$ decrease\\
\hline
1   & 2011-02-15 01:56                                                   & 11158 & X2.2      & CME (Halo)     & 2 & 2                    \\
2   & 2011-03-09 23:23                                                   & 11166 & X1.5      & CME           & - & -        \\
3  & 2011-09-06 22:20                                                   & 11283 & X2.1      & CME (Halo)    & - & -        \\
4  & 2011-09-07 22:38                                                   & 11283 & X1.8      & CME          & 1 & 1                  \\
5   & 2012-03-07 00:24                                                   & 11429 & X5.4      & CME (Halo)     & - & -        \\
6  & 2012-03-07 01:14                                                   & 11430 & X1.3      & CME (Halo)     & - & -        \\
7  & 2012-07-12 16:49                                                   & 11520 & X1.4      & CME (Halo)     &  1 & 0               \\
8  & 2013-11-08 04:26                                                   & 11890 & X1.1      & CME          & 1  & 0                  \\
9   & 2013-11-10 05:14                                                   & 11890 & X1.1      & CME          & - & -        \\
10   & 2014-01-07 18:32                                                   & 11944 & X1.2      & CME (Halo)     & 1  & 0                  \\
11  & 2014-03-29 17:48                                                   & 12017 & X1.0      & CME (Halo)     & 2  & 1                   \\
12 & 2014-09-10 17:33                                                   & 12158 & X1.6      & CME (Halo)    & 1 & 1                    \\
13 & 2014-10-22 14:06                                                   & 12192 & X1.6      & non-eruptive & - & -        \\
14 & 2014-10-24 21:15                                                   & 12192 & X3.1      & non-eruptive & - & -        \\
15 & 2014-10-25 17:08                                                   & 12192 & X1.0      & non-eruptive & - & -       \\
16 & 2014-10-26 10:56                                                   & 12192 & X2.0      & non-eruptive & - & -       \\
17  & 2014-11-07 17:26                                                   & 12205 & X1.6      & CME          & 1 &1                  \\
18  & 2014-12-20 00:28                                                   & 12242 & X1.8      & CME          & - & -       \\
19  & 2015-03-11 16:28                                                   & 12297 & X2.1      & CME          & 1 &1                     \\ 
\hline
\end{tabular}
\end{table*}

\section{Discussion}\label{sec:discussion_main}
\subsection{The evolution of the electruc current (density) at the flux-rope footpoint-areas}\label{sec:discussion_observation}
A first  observational study concerning four flares \citep{ChengDing2016} suggested that the electric current density decreases in the footpoints of  flux ropes.
This decrease can be related to straightening and untwisting of the magnetic field in the legs of the magnetic flux rope.
Using our 3D MHD simulation, we discussed how and why the electric current density evolves at the flux-rope footpoint-areas during flare eruption (Section~\ref{sec:dis_theory}).
Based on this simulation (Section~\ref{sec:dis_theory}), we found that $j_z$ and $I_z$ at the flux-rope footpoint-areas decrease by around 50\%-80\% during the impulsive phase of the solar flare.
This trend is consistent with the  previous observational study \citep{ChengDing2016}.
Using our 3D MHD simulation, we found  that the flux rope expansion causes a decrease of the electric current (density).
Based on the flux conservation for a cylindrical flux tube with a constant end-to-end twist, one can find that $j_z \propto L^{-1}$.
This relation is deeply discussed in Section~\ref{sec:dis_theory}.
During the flux rope expansion, the flux rope core is stretched that causes, firstly, untwist and then increases its length, which implies the decrease of the electric current density at the flux-rope footpoint-areas according to Eq.~\ref{eq:jz_l}.

However, \citep{ChengDing2016} noticed a significantly lower decrease of the electric current (density) than the decrease obtained in our simulation (7-13\% vs. 55-82\%, see Section~\ref{sec:dis_theory}).
To find an origin of the discrepancy between our simulation and this
observational study we analyzed  the temporal changes of the electric current (density) for a 
new sample of solar flares.

We studied all (19) X-class solar flares observed between 2011 and 2016 closely to the solar disc center, using EUV and UV images obtained by SDO.
In nine ROIs, we identified 11 flux-rope footpoint-areas.
In seven flux-rope footpoint-areas, we found a significant decrease (42\%-82\%) of the $j_z$ and $I_z$ during the eruptive phase of the solar flare.
In four flux-rope footpoint-areas, changes of $j_z$ and $I_z$ were smaller than the maximal error; thus, these cases were not considered in our further analysis.

Both our simulation and observations present a decrease of the electric current (density) with a similar amplitude at the flux-rope footpoint-areas during the impulsive phase of the solar flare.
This consistency suggests that our simulation can reproduce the same behavior as the observations.
Thus, the expansion of the flux rope is responsible for the decrease of the electric current (density) at the flux-rope footpoint-areas.

Our analysis of the 3D MHD simulation and observations suggests that the critical point of this study is the choice  of the region-of-interest that only covers the flux-rope footpoint-areas because the noisy region out of the flux-rope footpoint-areas can significantly reduce the value of the average electric current (density).
It was relatively easy to find the flux-rope footpoint-areas in our simulation using the magnetic field lines, but in the observations it was necessary to used more sophisticated methods.
In our opinion, in the previous observational work \citep{ChengDing2016}, the electric current (density) was computed in too large areas with a significant contribution from out of the flux-rope footpoint-areas.

We have now provided evidences of the evolution of the electric current (density) at the flux-rope footpoint-areas for many X-class flares, and in particular of its decrease.
Thus, can we better understand the $\bm{v};\bm{B}$ vs $\bm{E};\bm{j}$ paradigm issue studied the temporal evolution of the electric current (density) and magnetic field?
First, we found that the direct current is several times larger than return current at the flux-rope footpoint-areas; therefore, a net electric current exists.
However, the existence or absence of the net current cannot be used to solve the $\bm{v};\bm{B}$ vs $\bm{E};\bm{j}$ paradigm (see Section~\ref{sec:intro}).
Second, our simulation and observations showed that $\bm{\nabla \times B}$ diminishes during the flux rope expansion at the places where the flux rope is rooted.
The comparison of the characteristic time scale of the \bm{$j$} and \bm{$B$} evolution \citep{Vasyliunas2005} suggests that $\bm{\nabla \times B}$ diminishes is consistent with the $\bm{v};\bm{B}$ paradigm.
Thus, $\bm{v};\bm{B}$ are primary and $\bm{E};\bm{j}$ would be derived as their products.

\subsection{Limitations and prospects}\label{sec:discussion_limitations}
Our analysis is limited to a small sample of X-class solar flares  due to observational constrains e.g. observed between 2011 and 2016  in order to use SDO data; this period covers the maximum of the solar cycle 24.
Future studies should also examine the M-class solar flares which can concern  ARs with strong magnetic field concentrations. The sample can increase considerably because there is nearly 200 M-class flares per year when only 10 X-class flares per year.

Our study, as the first time, focuses on X-class solar flares and shows difficulties with the identification of 
the flux-rope footpoint-areas. For M-class flares it will be even worst but we should try and get a much larger sample for statistics. The identification of flux rope footpoints was really challenging.  Our methods are relatively empirical methods and need manual search with empirical assumptions. New methods could be derived to automatise the search.
We hope that further tests provided in the same way will be continued with new data, especially during the approaching solar maximum. 

Other limitations are related  directly to the magnetic field measurement.
We used only the photospheric magnetic field because the coronal magnetic field is significantly weaker, so difficult to measure.
Moreover, HMI vector magnetograms are obtained with a cadence of 12 minutes, and one can expect that some local processes on the electric current changes can happen with a shorter time-scale. The HMI vector magnetic field  is computed with five points in the  profiles, therefore the noise is quite important compared to the low sensibility of HMI.
Therefore, the new generation of space and ground-based high-resolution instruments and magnetographs are needed.
Nevertheless our conclusion is straightforward.  
%

\section{Conclusion}\label{sec:sumcon}

In this paper, we focused on the regions in the solar photosphere where the flux rope of the flare is rooted; this region is called the flux-rope footpoint-areas. 
In these areas, we studied the temporal evolution of the electric current (and electric current density) during eruptive flares.

We performed this analysis theoretically as well observationally. 
For the first time, we presented the evolution of the electric current (density) at the flux-rope footpoint-areas using the 3D MHD simulation.
In the simulation, selected magnetic field lines allowed us to identify where the flux rope is rooted.
Our simulation shows a decrease of direct $j_z$ and $I_z$ of 55-82\% at the flux-rope footpoint-areas.
%

Based on the SDO data, we identified the flux-rope footpoint-areas by defining the position of the ribbon hooks and the core of dimming regions.
For (semi)-bipolar active regions in which the identification of the ribbon hook or/and the dimming core was possible, we found eleven flux-rope footpoint-areas.
Seven of them present a decrease of the electric current (density) during the impulsive phase of the solar flare and after.
These decreases of electric current (density) of around  42\%-82\% at the flux-rope footpoint-areas are significantly larger than the maximal error.
In four flux-rope footpoint-areas, the electric current (density) does not present any clear trend of temporal evolution.
In summary, the decrease of the electric current (density) in the flare observations is consistent with our 3D simulation results.

Using our 3D model, we found that the temporal evolution of the direct electric current density in the area surrounded by the ribbon hooks shows a decreasing trend as $L^{-1}$. This fact  was predicted by analytic studies demonstrating that  electric current density ($j_z$) at the flux-rope footpoint-areas is inversely proportional to the flux rope length ($L$).
Thus, with our simulation we confirmed  that the decrease of the electric current density in the flux-rope footpoint-areas 
can be explained by flux rope expansion.

On the other hand we could give some arguments related to the so-called $\bm{v};\bm{B}$ paradigm 
applied to the solar corona.
The $\bm{v};\bm{B}$ paradigm defines that $\bm{v}$ is primary to $\bm{E}$ and $\bm{B}$ is primary to $\bm{j}$.
The decrease of the electric current (density) at the flux-rope footpoint-areas that we found in simulation and observations validates the theoretically predicted $\bm{v};\bm{B}$ paradigm of the solar atmosphere plasma.

\setcounter{table}{1}
\begin{table*}[h!]
\renewcommand{\thetable}{\arabic{table}}
\centering
\caption{The electric current and electric current density decrease at the flux-rope footpoint. The electric current (density) in ROIs (T1, T2) obtained from the simulation is dimensionless. In the ROIs obtained from observations (1A, 1B, 4A, 11A, 12A, 17A, 19A), the units of electric current (density) are presented in table. The values in a bracket present the percentage change of the electric current (density).} \label{tab:dd}
\begin{tabular}{cccccc}
\tablewidth{0pt}
\hline
\hline
    ROI & direct current & $\Delta j^\text{direct}_{z}$[$10^{-3} A\cdot m^{-2}$] (\%) & $ \Delta j^\text{net}_z$[$10^{-3} A\cdot m^{-2}$] (\%) & $\Delta I^\text{direct}_z$[$10^{11}$A](\%) & $ \Delta I^\text{net}_z$[$10^{11}$A](\%) \\
\hline
T1  & negative              & 1.8 (82.5)                  & 1.8 (82.5)                  & 0.9 (82.5)                  & 0.9 (82.5)                \\
T2  & positive              & 2.8 (55.4)                  &  2.8 (55.4)                  & 0.5 (55.4)                  & 0.5 (55.4)                \\ 
\hline
1A  & negative       & 22.3$\pm$3.2 (52.0)  & 28.3$\pm$9.1 (65.9)  & 3.9$\pm$0.5 (61.3)   & 4.2$\pm$1.4 (65.9) \\
1B  & positive       & 10.9$\pm$8.9 (58.1)  & 13.9$\pm$7.1 (56.4)  & 1.9$\pm$0.8 (82.1)   & 2.1$\pm$1.1 (96.4) \\
4A  & positive       & 24.9$\pm$24.4 (46.6) & 29.8$\pm$23.5 (69.4) & 4.0$\pm$2.9 (59.1)   & 4.4$\pm$3.6 (69.4) \\
11A & negative       & 32.9$\pm$3.0 (68.3)  & 31.9$\pm$15.1 (73.1) & 3.4$\pm$0.4 (71.1)   & 3.4$\pm$1.7 (73.1) \\
12A & negative       & 6.7$\pm$3.1 (42.5)   & 11.9$\pm$7.7 (83.9)  & 0.6$\pm$0.2 (65.1)   & 0.8$\pm$0.5 (83.9) \\
17A & negative       & 8.3$\pm$3.4 (65.4)   & 10.0$\pm$5.7 (90.6)  & 0.6$\pm$0.2 (77.4)   & 0.6$\pm$0.4 (90.6) \\
19A & positive       & 40.0$\pm$25.5 (42.2) & 42.4$\pm$26.2 (44.7) & 2.7$\pm$1.8 (44.5)   & 2.7$\pm$1.9 (44.7) \\
\hline
\end{tabular}
\end{table*}

\textbf{Acknowledgement}
This study benefited from financial support from the Programme National Soleil Terre (PNST) of the CNRS/INSU, as well as from the Programme des Investissements d'Avenir (PIA) supervised by the ANR. The work of KB is funded by the LabEx Plas@Par which is driven by Sorbonne Universit\'e. The numerical simulation used in this work was executed on the HPC center MesoPSL which is financed by the project Equip@Meso as well as the R\'egion Ile-de-France. This work has been done within the LABEX Plas@par project, and received financial state aid managed by the Agence Nationale de la Recherche, as part of the programme "Investissements d'avenir" under the reference ANR-11-IDEX-0004-02.

\bibliography{pp2}

\appendix

\section{Appendix information}\label{sec:appendix}
In Section~\ref{sec:AR11158}-~\ref{sec:AR12297}, we present three the most representative and diverse example of the flares with an identified flux-rope footpoint-areas.
Here, we discuss three additional active regions with an identified flux-rope footpoint-areas.

\subsection{AR 11283 }\label{sec:ar11283}
The active region NOAA AR 11283 generated 28 flares, between 31st August and 11th September 2011, two of them were X-class.
This active region was studied,  e.g. \citet{2013ApJ...771L..30J, 2014ApJ...780...55J, 2014ApJ...784..165R, 2014ApJ...787....7X,
2014ApJ...786...72Y,
2015ApJ...812..120R,
2016A&A...591A.141J} and 
\citet{2018AdSpR..61..673Y}. 

\begin{figure*}[ht!]
\epsscale{0.85}
\plotone{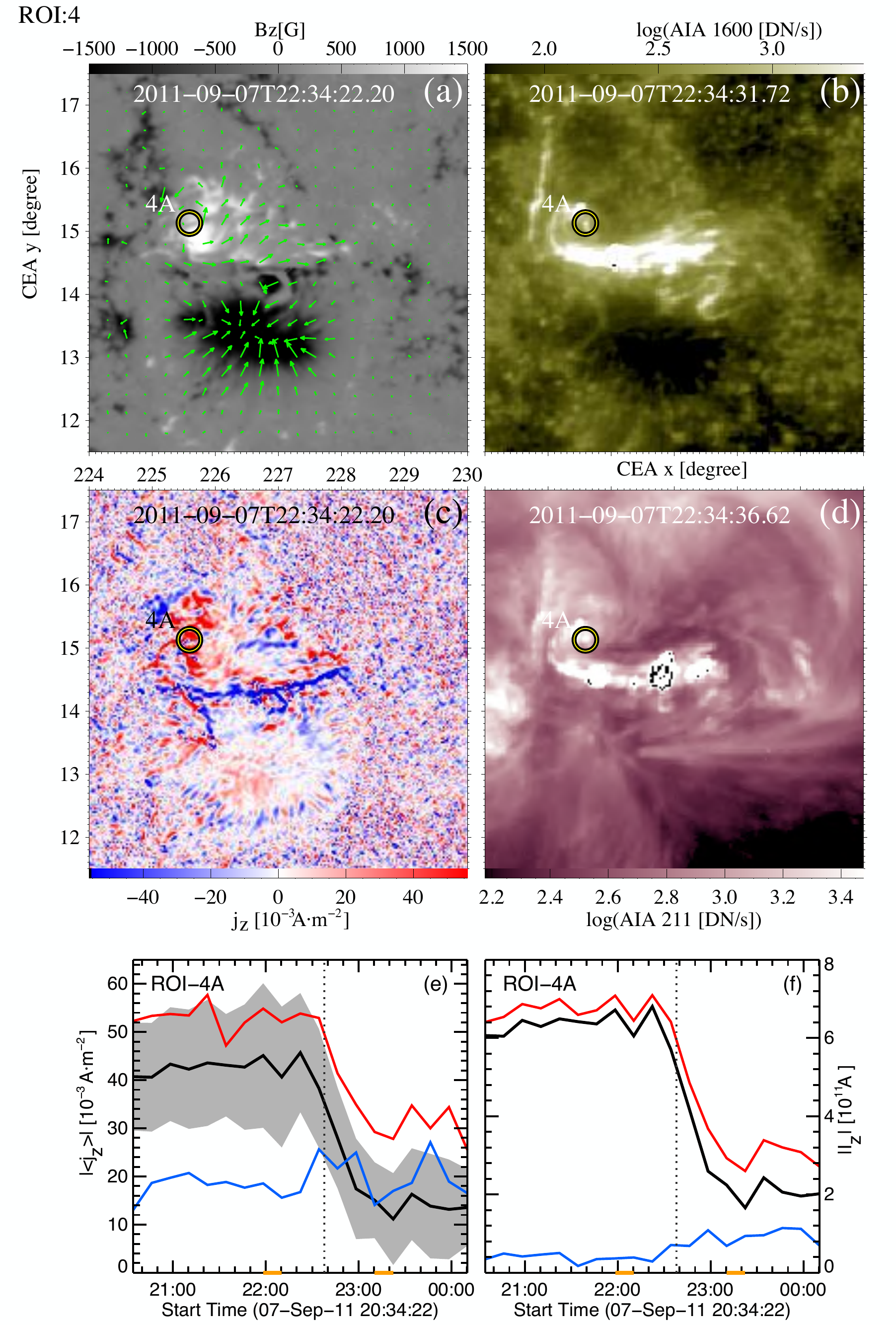}
\caption{
Context SDO images for the X1.8 class flare event area of 7th September 2011 (AR 11283) and temporal evolution of the electric current and electric current density at the footpoint of the flux rope; (a) HMI vector magnetogram, (b) AIA 1600\AA, (c) the vertical current density map, (d) AIA 211\AA, (e) temporal evolution of the electric current density at the footpoint of flux rope, (f) temporal evolution of the electric current at the footpoint of flux rope. The footpoints of the flux rope is marked by a yellow-black circle (ROI 4A). 
The photospheric magnetic field map (a) is presented in the same manner as in Figure~\ref{fig:roi1f}. Plots of the temporal evolution of the electric current density (e) and electric current (f) are presented in the same way as plots (a,c) and (b,d) in Figure~\ref{fig:roi1c}, respectively.
An animation of panels (a) - (d) is available. The video begins on September 7, 2011 at approximately 20:36:00 and ends the next day around 00:36:00. The
 realtime duration of the video is 48 seconds.
 \label{fig:roi4f}}
\end{figure*}

The active region AR 11283 had a bipolar magnetic field topology.
The negative magnetic field polarity is stronger than the positive one (Figure~\ref{fig:roi4f}a), but this asymmetry is weaker than in AR 12205.

The snapshot of the movie shows two ribbons in AIA 1600\AA~(Figure~\ref{fig:roi4f}b); the northern ribbon is wider than southern, and it is ended with a hook.
The southern ribbon presents a complex shape and evolves quickly; hence the identification of the flux-rope footpoint-areas was difficult.
We used auxiliary the AIA 211\AA~data (Figure~\ref{fig:roi4f}d) that shows a clear sigmoid that can be related to the flux rope.
However, we identified only one flux-rope footpoint-areas in this active region.

The direct $j_z$ and $I_z$ decrease during the impulsive phase of the flare, as shown in Figure~\ref{fig:roi4f}e and~\ref{fig:roi4f}f, respectively.
The same trend we noticed for the net $j_z$ and $I_z$.
The return $j_z$ and $I_z$ stay constant by the whole observation time.

\subsection{AR 12017}\label{sec:ar12017}
Between 23rd and 31st March 2014, the active region NOAA AR12017 was a source of 23 flares, including one X-class.
This X-class flare of 14th March 2014, was well-studied by many authors, e.g. \citet{2015ApJ...806....9K, 2016ApJ...827...38R, 2017SoPh..292...38W, 2018ApJ...860..163W}.

\begin{figure*}[ht!]
\epsscale{0.85}
\plotone{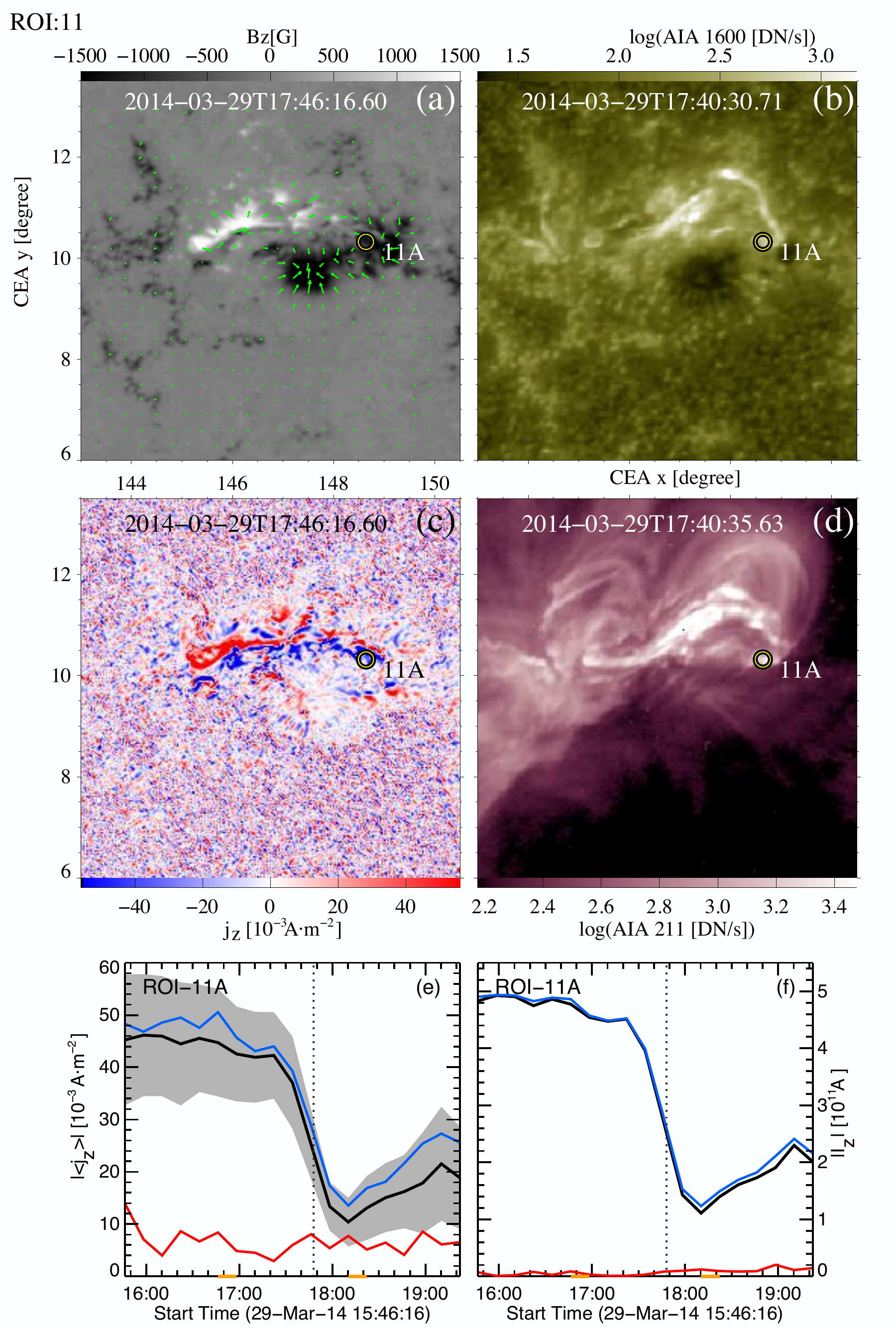}
\caption{
Context SDO images for the X1.0 class flare event area of 29th March 2014 (AR 12017) and temporal evolution of the electric current and electric current density at the footpoint of the flux rope; (a) HMI vector magnetogram, (b) AIA 1600\AA, (c) the vertical current density map, (d) AIA 211\AA, (e) temporal evolution of the electric current density at the footpoint of flux rope, (f) temporal evolution of the electric current at the footpoint of flux rope. The footpoints of the flux rope is marked by a yellow-black circle (ROI 11A). 
The photospheric magnetic field map (a) is presented in the same manner as in Figure~\ref{fig:roi1f}. Plots of the temporal evolution of the electric current density (e) and electric current (f) are presented in the same way as plots (a,c) and (b,d) in Figure~\ref{fig:roi1c}, respectively.
An animation of panels (a) - (d) is available. The video begins on March 29, 2014 at approximately 15:47:00 and ends the same day around 19:47:00. The
 realtime duration of the video is 48 seconds. \label{fig:roi11c}}
\end{figure*}

The active region AR 12017 has an asymmetric bipolar magnetic field topology (Figure~\ref{fig:roi11c}a) of two latitudinal elongated magnetic field concentrations.
The map of $j_z$ distribution (Figure~\ref{fig:roi11c}c) presents two electric current ribbons, but direct identification of the electric ribbon hook was problematic.
Two narrow ribbons are observed in AIA~1600\AA ~(Figure~\ref{fig:roi11c}b), but they evolve fast.
The identification of the single hook was only possible using additional images from the AIA~211\AA~channel (Figure~\ref{fig:roi11c}d), which shows the clear sigmoid and coronal core dimming region.
%

The plots of the temporal evolution of direct $j_z$ (Figure~\ref{fig:roi11c}e) and $I_z$ (Figure~\ref{fig:roi11c}f) at the flux-rope footpoint-areas show strong drops during the impulsive phase.
The same trend is noticed for the net $j_z$ and $I_z$.
The return $j_z$ and $I_z$ stay constant by the whole observation time. 

\subsection{AR 12158 }\label{sec:ar12158}
The active region NOAA 12158 produced 15 flares between 8th and 18th September 2014, including one X-class flare.
This X-class flare of 10th September 2014 was studied by many authors, e.g. \citet{2015ApJ...804...82C,
2015ApJ...807L..22G, 2015ApJ...811..139T, 2016ApJ...823...41D, 2017A&A...598A...3Z, 2017SoPh..292...11N, 2019A&A...621A..72A}.

\begin{figure*}[ht!]
\epsscale{0.85}
\plotone{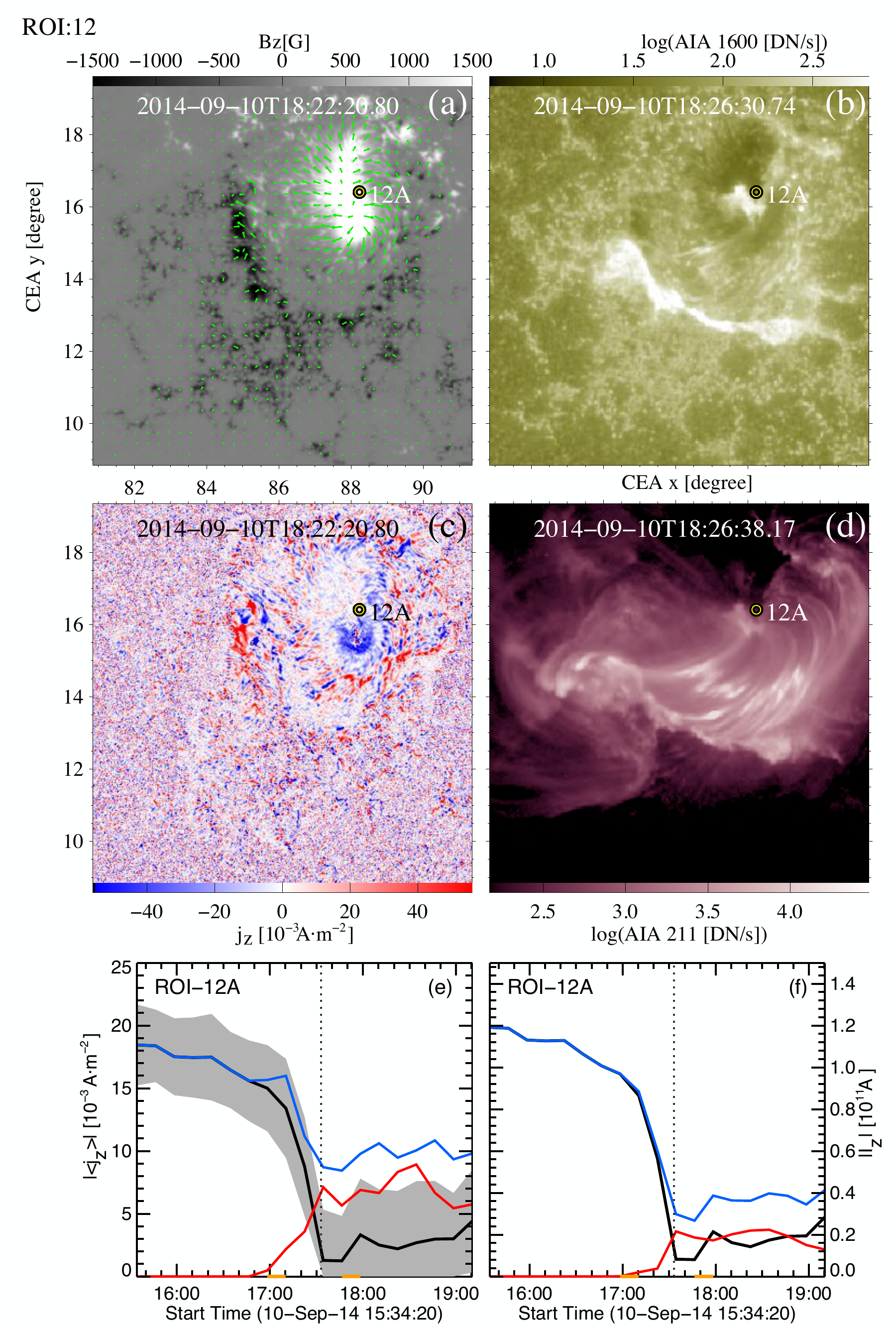}
\caption{Context SDO images for the X1.6 class flare event area of 10th September 2014 (AR 12158) and temporal evolution of the electric current and electric current density at the footpoint of the flux rope; (a) HMI vector magnetogram, (b) AIA 1600\AA, (c) the vertical current density map, (d) AIA 211\AA, (e) temporal evolution of the electric current density at the footpoint of flux rope, (f) temporal evolution of the electric current at the footpoint of flux rope. The footpoints of the flux rope is marked by a yellow-black circle (ROI 12A). 
The photospheric magnetic field map (a) is presented in the same manner as in Figure~\ref{fig:roi1f}. Plots of the temporal evolution of the electric current density (e) and electric current (f) are presented in the same way as plots (a,c) and (b,d) in Figure~\ref{fig:roi1c}, respectively.
An animation of panels (a) - (d) is available. The video begins on September 10, 2014 around 15:33:00 and ends the same day at approximately 19:21:00. The
 realtime duration of the video is 48 seconds.
 \label{fig:roi12c}}
\end{figure*}

The active region AR 12158 has a complex magnetic field topology (Figure~\ref{fig:roi12c}a) because the positive compact magnetic field concentration is surrounded by the arc of the negative magnetic patches from the south and positive magnetic patches from the north.
The AIA 1600\AA~(Figure~\ref{fig:roi12c}b) map of the active region shows two wide ribbons.
The ribbon in the north-east part of the active region is ended with a clear hook (ROI-12A).
This hook corresponds to the strong dimming region in AIA 211\AA~map (Figure~\ref{fig:roi12c}d) located in the area where the sigmoid is ended.
The electric current density map (Figure~\ref{fig:roi12c}c) does not allow for direct identification of the flux-rope footpoint-areas rather and presents several concentrated patches than elongated current ribbons such as in previous active region maps.

At the flux-rope footpoint-areas, the direct $j_z$ (Figure~\ref{fig:roi12c}e) and $I_z$ (Figure~\ref{fig:roi12c}f) decrease from around 20 minutes before the flare peak and continues this trend by the impulsive phase.
At the same time, return $j_z$ and $I_z$ increases.
This implies the sharp decrease of net $j_z$ and $I_z$.

\end{document}